\documentclass[10pt,letterpaper]{article}
\usepackage[T1]{fontenc}
\usepackage[utf8]{inputenc}
\usepackage{lmodern}
\usepackage[margin=1in]{geometry}
\usepackage{microtype}
\usepackage{amsmath,amssymb,amsthm}
\usepackage{graphicx,booktabs,xcolor}
\usepackage[numbers,sort&compress]{natbib}
\usepackage{algorithm,algpseudocode}
\usepackage{multirow,xspace,enumitem,tabularx,makecell,adjustbox}
\usepackage{listings,wrapfig,needspace,float,array,tcolorbox}
\usepackage{caption,subcaption,placeins}
\usepackage{commands}
\usepackage[hidelinks]{hyperref}
\usepackage{url}
\newtheorem{remark}{Remark}
\theoremstyle{plain}
\newtheorem{invariant}{Runtime Invariant}
\lstdefinestyle{prompt}{
  basicstyle=\ttfamily\fontsize{8.5}{10.2}\selectfont,
  columns=fullflexible,
  keepspaces=true,
  breaklines=true,
  breakatwhitespace=true,
  showstringspaces=false,
  frame=single,
  rulecolor=\color{black!30},
  backgroundcolor=\color{black!2},
  framesep=5pt,
  aboveskip=6pt,
  belowskip=6pt,
  upquote=true
}
\lstdefinestyle{sql}{
  style=prompt,
  language=SQL,
  basicstyle=\ttfamily\small,
  backgroundcolor=\color{white},
  keywordstyle=\bfseries
}

\title{\textbf{Post-SQLFix: Property-Guided Repair of\\LLM-Generated SQL with Runtime Verification}}
\author{
\begin{tabular}{c}
Zhihao Guo\textsuperscript{1}\thanks{These authors contributed equally to this work.}
\quad
Zonghan Wu\textsuperscript{2}\footnotemark[1]
\quad
Yilei Shao\textsuperscript{2} \quad
Haizhou Du\textsuperscript{3} \quad
Nan Tang\textsuperscript{4}
\\[3pt]
Xiao Cheng\textsuperscript{5}\thanks{Corresponding authors: Xiao Cheng and Huan Huo.}
\quad
Huan Huo\textsuperscript{1}\footnotemark[2]
\\[9pt]
\small \textsuperscript{1}University of Technology Sydney
\small \textsuperscript{2}East China Normal University
\\
\small \textsuperscript{3}Shanghai University of Electric Power
\small \textsuperscript{4}Hong Kong University of Science and Technology (GZ)
\\
\small \textsuperscript{5}Macquarie University
\end{tabular}
}
\date{}
\hypersetup{
pdftitle={Post-SQLFix: Property-Guided Repair of LLM-Generated SQL with Runtime Verification},
pdfauthor={Zhihao Guo, Zonghan Wu, Yilei Shao, Haizhou Du, Nan Tang, Xiao Cheng, Huan Huo},
pdfsubject={Text-to-SQL repair}
}
\begin{document}
\maketitle
\begin{abstract}
Large Language Models (LLMs) for Text-to-SQL frequently
generate queries with \blue{syntax errors, invalid identifier
references, or violations of SQL scoping and typing rules}. Repairing these errors is a program correctness problem, yet existing approaches rely on Execution Feedback (EF), whose symptom-level, dialect-specific error messages force the LLM to speculate on root causes. We identify an {Arbitration Paradox} in this paradigm: \blue{the system relies on unverified models both to propose repairs and to make the final judgment on whether those repairs satisfy the required properties.}
We resolve this paradox by reformulating Text-to-SQL repair as property-guided program repair. Our framework, \thistool, canonicalizes SQL into a dialect-aware Canonical Query Structure (CQS) and validates it through a hierarchical static analyzer with three \blue{stratified} property layers (Syntactic, Context-Free Structural, Context-Sensitive Semantic). The analyzer produces precise, schema-aware violation diagnostics \blue{to guide the LLM's repair proposals. A Verify-Before-Apply Temporal Monitor then accepts only repairs that reduce the violation count, do not reintroduce resolved violations, and satisfy the operator contracts. These runtime invariants are bounded by the declared rule family and operator contracts, while a finite retry budget ensures halting.}
On three established benchmarks (BIRD, Spider, Mini-Dev) 
spanning over 3,000 evaluation instances, three SQL dialects 
(SQLite, MySQL, PostgreSQL), and \blue{six model configurations}, 
\thistool\ resolves 77.8\% of the execution errors that 
EF leaves unrepaired, while requiring \blue{less than half as many repair
iterations} and 30\% fewer tokens. Crucially, \blue{accepted transitions under the monitor 
introduce no statically detectable structural regressions in our evaluation, translating 
checked repair fidelity }
into absolute accuracy improvements of up to +11.6 \blue{percentage points}.
\end{abstract}

\section{Introduction}
\label{sec:intro}

Large language models (LLMs) can translate natural-language questions into executable SQL~\cite{SPIDER, wang2019ratsql, hwang2019sqlova}, yet the resulting queries frequently \blue{contain} syntactic flaws, schema-entity type mismatches, and structural scoping errors~\citep{SPIDER, wang2019ratsql}. Unlike compiled languages, SQL offers no static type-checking compilation step between generation and execution: an ill-formed query is discovered only when the database engine rejects it at runtime or, worse, when it silently returns incorrect results. Repairing these errors is therefore a {program correctness} problem, orthogonal to the choice of generator: any LLM-based Text-to-SQL pipeline can benefit from a principled post-generation repair layer. Such a layer demands the kind of formal intermediate representations, property specifications, and bounded \blue{runtime checks} that are central to the programming-languages tradition.

The dominant repair paradigm relies exclusively on the database engine for runtime verification.
\textbf{Execution-Feedback (EF)} pipelines~\citep{sun2023sqlpalm, deng2024sqlcraft, chen2023selfdebug} run the candidate query, capture the engine's error message, and prompt the LLM to fix it. Although these runtime signals are faithful, they are also {symptom-level}: messages such as \code{Syntax error near WHERE} lack schema- and intent-level context, vary across dialects, and conflate multiple root causes into a single opaque trace. More fundamentally, EF provides {no signal at all} for queries that execute successfully yet return wrong results (e.g., scoping violations, invalid aggregations, or empty-result cases). The LLM must still diagnose root causes and synthesize repairs based on the ambiguous evidence from EF, routinely introducing new errors, a well-documented fix-and-regress cycle~\citep{huang2024largelanguagemodelsselfcorrect, kim2024llm}.

\begin{figure*}[t]
    \centering
    \includegraphics[width=0.95\textwidth]{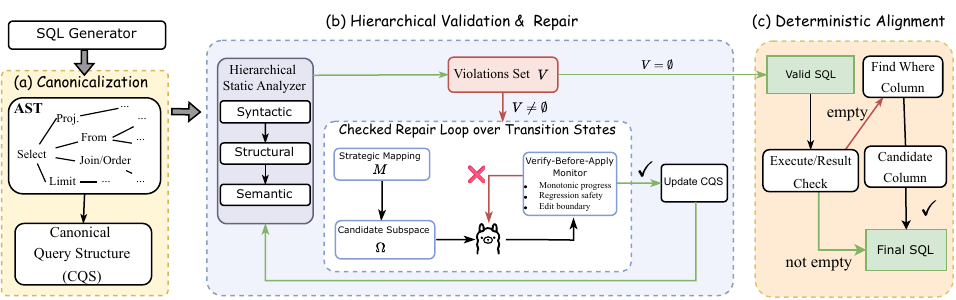}
    
    \caption{Overview of \thistool. (a)~Canonicalization into CQS. (b)~\blue{Hierarchical Validation \& Repair}. (c)~\blue{\detalign} for empty-result queries.}
    \label{fig:archi_design}
    
\end{figure*}

This failure is not incidental. It reflects a structural limitation we term the \textbf{Arbitration Paradox}: \blue{the system relies on unverified models both to propose repairs and to make the final judgment on whether those repairs satisfy the required properties.} Whether the architecture is self-correction~\citep{dinsql, dailsql, madaan2023selfrefine}, retrieval-augmented generation~\citep{chesssql, gensql, RESDSQL}, or multi-agent coordination~\citep{macsql, magic, SQLFixAgent, agent_fail}, \blue{successful execution alone does not verify that a repair preserves previously correct query structure. If the final judgment of repair compliance is delegated to probabilistic LLM reasoning}~\citep{gu2025surveyllmasajudge}, \blue{the unverified mechanism proposing the repair also becomes the authority on whether it should be accepted.}

The paradox exposes two concrete gaps. First, a \textbf{diagnostic gap}: no deterministic mechanism exists to lift the ambiguous, engine-specific runtime symptoms of EF pipelines into precise, schema-aware property violations; the LLM must map an opaque error trace, or for silently wrong queries \emph{no} trace at all, to a structured root cause, and does so unreliably. Second, a \textbf{repair gap}: even given an accurate diagnosis, no principled mechanism constrains the repair search space or enforces monotonic progress; existing pipelines \blue{check} neither that a proposed fix preserves previously correct structure ({soundness}) nor that the process converges within bounded iterations ({termination}), and the LLM may freely introduce new violations while patching old ones.

A natural response is to turn to existing SQL static analysis. However, industry linters such as SQLFluff~\citep{sqlfluff} enforce only syntactic formatting rules without schema-aware semantic validation. Prior work on statically analyzing dynamically generated SQL~\citep{gould2004static} targets human-written application code and does not account for the distinctive error distribution of LLM generators (e.g., hallucinated table names, foreign-key-agnostic joins). Formal equivalence checkers~\citep{wang2024verieql} require a ground-truth reference query, which is unavailable at inference time. All of these tools are {passive validators}: they may detect errors but provide no mechanism to synthesize repairs or to \blue{check} that a proposed fix does not introduce new violations. To bridge this gap, we draw inspiration from the broader programming languages community, where property-guided program repair~\cite{bugariu2022automatically, ProveNFix} has proven highly successful in bypassing ambiguous runtime errors to synthesize \blue{checked} fixes. The efficacy of these formal methods in traditional software development suggests that a similar paradigm is highly viable for the Text-to-SQL problem.

\vspace{1mm}
\emph{Our Solution.}
We resolve the Arbitration Paradox by reformulating Text-to-SQL repair as a \textbf{property-guided program repair} problem~\citep{ProveNFix, bugariu2022automatically}, replacing speculative execution feedback with verifiable static analysis and formally bounded synthesis. Our framework, \thistool, is a {generator-agnostic} post-processing layer: it takes the SQL output of any existing Text-to-SQL pipeline and strengthens it through formal analysis and repair. The direct comparison target is the EF repair paradigm that such pipelines currently rely on.


To close the repair gap, we formalize the repair process as a state-transition system governed by a \textbf{\blue{verify-before-apply} temporal monitor}. For each detected violation, the monitor first computes a bounded candidate subspace from the schema and dialect configuration, constraining the LLM to propose patches within that subspace. A candidate transition is accepted only if the violation set strictly decreases; otherwise it is rejected and the LLM is re-prompted. \blue{This monotonicity invariant directly checks bounded soundness (\blue{no previously resolved violation recurs}) and {bounded halting} (the loop terminates under the finite retry budget, returning either a zero-violation CQS or the last accepted state).}


Figure~\ref{fig:archi_design} illustrates the end-to-end pipeline. (a)~The raw SQL is parsed into a dialect-aware Abstract Syntax Tree and lifted into a Canonical Query Structure~(CQS), a normalized intermediate representation that decouples relational semantics from dialect-specific surface syntax. (b)~\blue{The hierarchical validation and repair stage combines the static analyzer with a checked repair loop. The analyzer reports violations of the declared $\alpha/\beta/\gamma$ rules, closing the diagnostic gap. These violations guide candidate construction, LLM repair proposals, and validation before updating the CQS.} (c)~Queries that pass all static layers but return empty results are routed to a data-driven alignment stage that conservatively adjusts filter predicates using empirical evidence from the database. We illustrate this paradox and our workflow through a concrete query repair in \S~\ref{sec:overview}.
\thistool\ is the first framework to provide \blue{\emph{formally specified runtime invariant checks}} for LLM-generated SQL repair. Our contributions are summarized as follows:
\begin{itemize}[nosep, leftmargin=*]
\item A \textbf{property-guided formulation} of Text-to-SQL repair, grounded in a dialect-aware intermediate representation (CQS) and a \blue{stratified} static analyzer that replaces ambiguous execution feedback with precise, schema-aware property violations (\S\ref{sec:formal_framework}).
\item A \textbf{\blue{verify-before-apply} temporal monitor} with \blue{runtime-enforced monotonicity}: every accepted repair transition \blue{strictly decreases the violation count and satisfies the history and preservation checks} (\blue{Runtime Invariant}~\ref{thm:repair-soundness-main})\blue{. A finite retry budget also bounds rejected proposals} (\S\ref{sec:repair_formalism}).
\item An \textbf{empirical evaluation} across three SQL dialects (SQLite, MySQL, PostgreSQL) and \blue{six models} showing up to \textbf{+11.6 \blue{percentage points}} of execution accuracy improvement, over 58\% fewer repair iterations, and a 100\% interception rate of LLM-induced regressions (\S\ref{sec:experiments}).
\end{itemize}

\section{Overview: A Motivating Example}
\label{sec:overview}

We walk through a concrete query repair to show how \thistool\ works end-to-end. The example previews the formal mechanisms of \S\ref{sec:formal_framework}--\S\ref{sec:repair_formalism} and exposes failure modes of the Execution-Feedback (EF) paradigm that motivate our design.

\subsection{Problem Setup}
Consider a natural-language question posed to a Text-to-SQL system over a relational financial database:
\begin{quote}
\textit{``What is the district ID for the account associated with order 33333?''}
\end{quote}

\noindent An upstream LLM generator~\cite{tasql,gensql,openai2024gpt4technicalreport} produces the following candidate:

\begin{lstlisting}[language=SQL, basicstyle=\ttfamily\small]
SELECT account.district_id
FROM order
WHERE order_id = 33333;
\end{lstlisting}

Two errors lurk in this query, each from a different semantic layer: (1)~\code{order} is a SQL reserved keyword, so \blue{its unescaped use causes a syntax error}; (2)~\code{account.district\_id} references a table absent from the \code{FROM} clause, a column-resolution failure that only surfaces once the first error is fixed.

Under EF, both errors produce undifferentiated engine messages. The LLM must disentangle them from ambiguous runtime output, and routinely fails: the EF pipeline takes \textbf{three repair iterations} (4{,}217 tokens), each round fixing one error while reintroducing the other. \thistool\ resolves both in a \textbf{single accepted transition} (799 tokens) by decomposing them across distinct analysis layers. We also use this example to show how the \blue{Verify-Before-Apply Temporal Monitor} catches a silent regression that EF would miss entirely.

\subsection{Canonicalization into CQS}
\blue{After deterministic keyword escaping,} \thistool\ parses the input into a dialect-aware AST and lifts it into a \textbf{Canonical Query Structure}~(CQS), the intermediate representation that all subsequent analysis operates on. The resulting structure~$\Phi$ records:
\begin{align*}
\Phi.\Pi       &= \{ \texttt{account.district\_id} \}  && \text{(projection set)} \\
\Phi.\rho.T    &= \{ \texttt{order} \}                  && \text{(relation scope)} \\
\Phi.\rho.J    &= \emptyset                              && \text{(join set)} \\
\Phi.\rho.F    &= \{ \texttt{order\_id = 33333} \}       && \text{(filter predicates)}
\end{align*}
The CQS strips away dialect-specific syntax (quoting conventions, reserved-keyword sets, function aliases) while preserving the relational structure that diagnosis needs. Because the representation is dialect-neutral, the \blue{analyzer is dialect-parametric: adding a dialect requires a fixed configuration rather than changes to the CQS rules} (\S\ref{sec:canonicalization}).

\subsection{Hierarchical Static Analysis}
The analyzer validates~$\Phi$ against the target schema~$\Sigma$ through three \blue{stratified} property layers (Algorithm~\ref{alg:analyzer}). Each layer targets a distinct error class; evaluating them in \blue{dependency} order avoids the tangled, all-at-once error reports that EF produces.

\paragraph{Layer $\alpha$ (Syntactic).}
The parser identifies \code{order} as a SQL reserved keyword used as a bare table identifier. The $\alpha$-layer applies a \textbf{deterministic} rewrite: it walks the AST \code{Identifier} nodes, checks membership in the dialect's reserved-keyword set, and emits the escaped form \code{`order`}. No LLM is involved; the syntactic violation is resolved before any neural component runs.

\paragraph{Layer $\beta$ (Context-Free Structural).}
Clause dependencies are well-formed: no aggregation-scoping or cross-clause reference violations arise.

\paragraph{Layer $\gamma$ (Context-Sensitive Semantic).}
The Identifier Binding check iterates over every attribute in~$\Phi$. For the projected column \code{account.district\_id}, it tests whether any table in the current relation scope declares a column \code{district\_id}:
\[
\neg\,\exists\, t \in \Phi.\rho.T \;\text{s.t.}\; (t, \texttt{district\_id}) \in \Sigma.\mathcal{C}
\]
The predicate fails: no relation in scope has this column. A schema-wide search locates it in table \code{account}, which is absent from~$\Phi.\rho.T$. The analyzer emits:
\begin{equation*}
v_1 = v_{\text{bind}}(\texttt{account.district\_id},\; \text{layer}=\gamma,\; \text{found\_in}=\texttt{account})
\end{equation*}
That is, $v_1$ records the unresolved column, the layer that flagged it, and the schema table where the column actually resides, giving the repair engine a precise, actionable diagnostic.

\smallskip
\noindent\textit{Comparison with EF.}
At this point \thistool\ has a complete, layered diagnosis without having executed the query. An EF pipeline would return only the first engine error (the parse failure on \code{order}). The column-resolution problem stays hidden until the keyword error is fixed, so the developer (or the LLM) discovers errors one at a time across multiple round trips.

\subsection{Constrained Repair Synthesis}
\label{sec:overview_synthesis}
Since $V \neq \emptyset$, the Plan Synthesis Engine (Algorithm~\ref{alg:synthesis}) is invoked. The strategic mapping dispatches~$v_1$ to the \code{add\_join} operator, triggering a \textbf{Foreign-Key~(FK) Graph Traversal}:

\begin{enumerate}[nosep]
    \item \textbf{Connectivity analysis.} A graph over~\blue{$\Phi.\rho.T\cup\{\texttt{account}\}$, where \code{account} comes from the binding diagnostic,} is built from the existing join predicates. Breadth-first search finds two disconnected components, $\{\texttt{`order`}\}$ and $\{\texttt{account}\}$, with no join bridging them.
    \item \textbf{FK path search.} Depth-first search over the foreign-key graph $(\Sigma.\mathcal{T}, \Sigma.\mathcal{K}_{\text{fk}})$ discovers:\\ $\texttt{`order`.account\_id} \xrightarrow{\text{FK}} \texttt{account.account\_id}$.
    \item \textbf{Candidate generation.} The FK path is compiled into a \code{JOIN} predicate. Only tables not already in~$\Phi.\rho.T$ are introduced.
\end{enumerate}

\noindent The engine yields a bounded \textbf{Candidate Subspace}:
\[
\Omega_{\text{join}}(\blue{\{\texttt{order}, \texttt{account}\}}) = \{\omega_{\text{add\_join}}(\texttt{account}, \texttt{`order`.account\_id = account.account\_id})\}
\]
The deterministic $\alpha$-escape and the FK-derived join candidate for~$v_1$ are compiled into a constrained prompt that narrows the LLM's task from open-ended speculation to selection within a verified subspace.

\subsection{\blue{Verify-Before-Apply Temporal Monitor}: Regression Interception}
The LLM receives the specification but \textit{over-corrects}: it inserts the recommended join yet silently drops the \code{WHERE} clause:

\begin{lstlisting}[language=SQL, basicstyle=\ttfamily\small]
SELECT account.district_id
FROM `order`
JOIN account
    ON `order`.account_id = account.account_id;
\end{lstlisting}

Under EF, this query runs without error, yet it returns the district ID for \textit{every} account rather than the one associated with order~33333. EF has no way to detect this regression without an external oracle.

The \blue{monitor} re-canonicalizes the output into~$\Phi'$ and checks the transition against three structural invariants. The original violation~$v_1$ is gone, but \textbf{Invariant Preservation} flags an unauthorized mutation:
$$
\Phi.\rho.F = \{ \texttt{order\_id = 33333} \},\quad \Phi'.\rho.F = \emptyset
$$
The filter set was not \blue{authorized by any admissible repair plan}, yet the LLM removed it. The transition is \textbf{rejected}, and the LLM is re-prompted with an explicit preservation constraint.

\subsection{Convergence: Accepted Repair}
On re-prompting, the LLM produces a repair that retains every existing clause:

\begin{lstlisting}[language=SQL, basicstyle=\ttfamily\small]
SELECT account.district_id
FROM `order`
JOIN account
    ON `order`.account_id = account.account_id
WHERE `order`.order_id = 33333;
\end{lstlisting}

The Monitor re-validates: $V(\Phi'', \Sigma, \Delta) = \emptyset$. The reserved keyword is escaped, the join follows the FK-derived path, every column resolves against its schema-declared table, and $\Phi''.\rho.F \supseteq \Phi.\rho.F$. The transition is \textbf{accepted}, and the repair completes without any database execution having occurred.

\medskip
\noindent This walkthrough previews three properties formalized in \S\ref{sec:formal_framework}--\S\ref{sec:repair_formalism}:
\begin{itemize}[nosep]
\item \textit{Regression blocking}: the Monitor rejected a syntactically valid but structurally mutated candidate, catching a silent regression that EF would have passed through (\S\ref{sec:soundness}).
\item \blue{\textit{Bounded halting}}: repair converged in two monitor rounds rather than spiralling through a fix-and-regress cycle; the monotonicity invariant caps accepted transitions at~$|V_0|$ (\S\ref{sec:repair_formalism}).
\item \textit{\blue{Dialect-parametricity}}: the entire $\alpha\beta\gamma$ analysis ran on the CQS with zero database executions, requiring only the \blue{fixed} dialect configuration~$\Delta$ (Remark~\ref{rem:dialect_cost}).
\end{itemize}

\section{Formal Framework: Property Specification via Analysis}
\label{sec:formal_framework}

This section details the formal foundations of \thistool, designed to address the Arbitration Paradox. We bridge the neuro-symbolic divide by replacing speculative Execution Feedback (EF) with strict, property-guided static analysis. As shown in Figure~\ref{fig:archi_design}, our pipeline operates by lifting the raw SQL query into a Canonical Query Structure~(CQS), upon which a hierarchical static analyzer enforces strict structural and semantic properties.

\begin{table*}[!t]
\centering
\footnotesize
\begin{tabular}{@{}p{0.35\textwidth}|p{\dimexpr0.65\textwidth-2\tabcolsep-\arrayrulewidth\relax}@{}}
\toprule
\textbf{Validation Layer} & \textbf{Error Categories} \\
\midrule
\textbf{Syntactic ($\alpha$)} & 
This layer handles precursors to semantic analysis, such as typos \\
\midrule
\textbf{Context-Free Structural ($\beta$)} & 
\textbf{Clause-related:} Clause Redundancy \\
& \textbf{Function-related:} Aggregate Functions \\
& \textbf{Operator-related:} Logical Operator (precedence issues) \\
& \textbf{Other:} ASC/DESC, DISTINCT (structural usage) \\
\midrule
\textbf{Context-Sensitive Semantic ($\gamma$)} & 
\textbf{Attribute-related:} Attribute Mismatch/Missing/Redundancy \\
& \textbf{Table-related:} Table Mismatch/Missing/Redundancy \\
& \textbf{Join-related:} Join Condition Mismatch, Join Type Mismatch \\
& \textbf{Function-related:} Date/Time, Conversion, Math, String, Conditional (type/schema dependent usage) \\
& \textbf{Value-related:} Data Format Mismatch \\
\midrule
\textbf{Empty Result Warnings} & 
\textbf{Condition-related:} Explicit Condition Mismatch/Missing \\
& \textbf{Value-related:} Value Mismatch \\
& \textbf{Subquery-related:} Subquery Mismatch \\
\bottomrule
\end{tabular}
\caption{Hierarchical Taxonomy of SQL Validation Errors}
\label{tab:validation_taxonomy}

\end{table*}

\subsection{Abstracting the Domain}
\label{sec:canonicalization}
To establish a rigorous static analysis environment, we first formalize the target database schema as a 5-tuple context:
\begin{equation}
    \Sigma = (\mathcal{T}, \mathcal{C}, \mathcal{K}_{\text{pk}}, \mathcal{K}_{\text{fk}}, \Theta),
\end{equation}
where $\mathcal{T}$ is a set of base relation names and $\mathcal{C}$ is the set of all fully qualified columns. The sets $\mathcal{K}_{\text{pk}} \subseteq \mathcal{C}$ and $\mathcal{K}_{\text{fk}} \subseteq \mathcal{C} \times \mathcal{C}$ capture primary and foreign key integrity constraints, respectively. Finally, $\Theta: \mathcal{C} \rightarrow \text{Type}$ maps each column to its strict data type signature.
While $\Sigma$ abstracts the static schema environment, raw SQL instances exhibit significant syntactic variance across execution engines (e.g., SQLite, PostgreSQL). To perform sound property verification, we must decouple the query's core relational semantics from its dialect-specific surface syntax. We \blue{parse} the input string $Q_{\text{str}}$ into a dialect-aware Abstract Syntax Tree (AST)\blue{, then lift the AST into} the \textbf{Canonical Query Structure (CQS)}, denoted $\Phi$. \blue{Recoverable $\alpha$ diagnostics such as reserved-keyword escaping are normalized by deterministic $\alpha$-operators before $\beta/\gamma$ analysis; if the parser returns $\bot_\pi$ rather than a CQS, Algorithm~\ref{alg:analyzer} emits an $\alpha$-layer violation and does not pass a malformed query to deeper layers. Only successfully parsed queries proceed to $\beta/\gamma$ analysis, where ambiguous or unresolved identifiers are handled by canonical binding checks.} Despite divergences in lexical symbols or grammatical constructs among SQL dialects, \blue{the CQS records the query's relational dependencies in a uniform representation}. By normalizing the query representation \blue{after parsing}, CQS substantially reduces the analytical discrepancies induced by dialect variance, providing a stable foundation for schema-independent structural verification (Layer $\beta$, Section~\ref{sec:validation_engine}) while safely delegating execution-specific constraints to a decoupled dialect configuration $\Delta$ (Remark~\ref{rem:dialect_cost}).

\begin{definition}[Core SQL Fragment $\mathcal{L}_{\text{sql}}$]
\blue{The CQS extraction targets the SQL fragment $\mathcal{L}_{\text{sql}}$ with the following abstract query structure:}
\begin{align*}
    \mathit{Query} &::= \blue{\text{[WITH }\mathit{CTEs}\text{] }}\text{SELECT }\blue{\text{[DISTINCT] }}\mathit{Proj} \text{ FROM } \mathit{Rel} \\
    &\quad\;\; \text{[WHERE }\mathit{Cond}\text{]} \text{ [GROUP BY }\mathit{Cols}\text{]} \text{ [HAVING }\mathit{Cond}\text{]} \\
    &\quad\;\; \text{[ORDER BY }\mathit{OrdCols}\text{]} \text{ [LIMIT }\mathit{Num}\text{]} \\
    \blue{\mathit{CTEs}} &\blue{::= \mathit{Alias}\text{ AS (}\mathit{Query}\text{)} \mid \mathit{Alias}\text{ AS (}\mathit{Query}\text{), }\mathit{CTEs}} \\
    \mathit{Proj} &::= \mathit{Expr}\blue{\text{ [AS }\mathit{Alias}\text{]}} \mid \mathit{Expr}\blue{\text{ [AS }\mathit{Alias}\text{]}}\text{, }\mathit{Proj} \\
    \mathit{Rel} &::= \mathit{Table} \mid \mathit{Rel} \text{ JOIN } \blue{\mathit{Rel}} \text{ ON } \mathit{Cond} \mid \text{(} \mathit{Query} \text{)} \text{ AS } \mathit{Alias}
\end{align*}
\blue{Here, $\mathit{Expr}$ denotes a SQL expression, including aggregate expressions, and $\mathit{Cond}$ denotes a condition used in \code{WHERE},
\code{HAVING}, or \code{JOIN ON}. The fragment includes non-recursive common table expressions (CTEs) and excludes window constructs.}
\end{definition}

\begin{definition}[Canonical Query Structure]
The CQS $\Phi$ distills a query's essential operational logic into a 6-tuple intermediate representation:
\begin{equation}
\Phi = (\Pi, \rho, \Gamma, \Omega_{\text{ord}}, A, S),
\end{equation}
\begin{itemize}[leftmargin=*, nosep]
    \item $\Pi$: \textbf{Projection Specification}, a set of projected \blue{expressions}.
    \item $\rho$: \textbf{Data Source Specification}, a tuple $(T, J, F)$ defining \blue{source relations $T$ (base tables and derived sources)}, join predicates $J$, and filter conditions $F$.
    \item $\Gamma$: \textbf{Aggregation Specification}, a tuple $(G, H)$ defining grouping attributes and \code{HAVING} conditions.
    \item $\Omega_{\text{ord}}$: \textbf{Presentation Specification}, capturing ordering\blue{, cardinality limits, and \code{DISTINCT} modifier}.
    \item $A$: \textbf{Alias Assignments}, mapping relations or attributes for identifier resolution.
    \item $S$: \textbf{Nested Scopes}, \blue{containing non-recursive CTEs and subqueries}. Each scope is defined as an independent CQS $\Phi'$, enabling modular, recursive static analysis of arbitrary nesting.
\end{itemize}
\end{definition}


\subsection{Property Extraction and Hierarchical Validation}
\label{sec:validation_engine}

The core of our property-guided repair is the static analyzer, which inspects the CQS $\Phi$ against a \blue{stratified} hierarchy of integrity properties. As summarized in Table~\ref{tab:validation_taxonomy}, we use the Text-to-SQL error classifications~\citep{NL2SQL-BUGs} \blue{to organize the rule family implemented by \thistool. The analyzer covers the statically checkable structural and schema-dependent subset; value-level failures are handled only when explicit data evidence is available, as in \detalign\ (\S\ref{sec:delta_layer}).} Furthermore, we extend beyond standard empirical categorizations by incorporating rich, fault-tolerant parsing checks to capture early-stage generation anomalies. Structured as a \blue{\textbf{stratified dependency order} $\alpha \prec_{\mathrm{sd}} \beta \prec_{\mathrm{sd}} \gamma$, the layers are evaluated in a dependency-respecting sequence: $\beta$ requires a parseable CQS from $\alpha$, and $\gamma$ requires stabilized scope, alias, and export information from $\beta$. This operational stratification determines validation and repair priority without imposing an abstract order over arbitrary rules.} Foundational violations are therefore addressed before context-sensitive constraints, enabling the analyzer to report all statically detectable violations \blue{for a parsed CQS} in a single pass. Algorithm~\ref{alg:analyzer} formalizes this extraction process. To state validation rules concisely, let $\mathcal{O}(\Phi)$ denote the set of all operator applications (e.g., aggregations, type casts, arithmetic functions) in~$\Phi$.

\begin{algorithm}[tbp]
\caption{Hierarchical Property Analysis ($\alpha \prec_{\blue{\mathrm{sd}}} \beta \prec_{\blue{\mathrm{sd}}} \gamma$)}
\label{alg:analyzer}
\footnotesize
\begin{algorithmic}[1]
\Require \blue{SQL string $Q_{\text{str}}$, Schema $\Sigma$}, Dialect $\Delta$
\Ensure \blue{Canonical} violation set $V$

\State $V \gets \emptyset$

\Statex \Comment{\textbf{Layer $\alpha$: Syntactic Well-Formedness}}
\State \blue{$Q_{\text{str}} \gets \textsc{AlphaNormalize}(Q_{\text{str}}, \Delta)$} \Comment{\blue{Recoverable keyword/quote normalization}}
\State $\Phi \gets \pi(Q_{\text{str}}, \Delta)$ \Comment{\blue{Lift SQL into CQS; return $\bot_\pi$ on parse failure}}
\If{\blue{$\Phi = \bot_\pi$}}
    \State \blue{$\kappa \gets \textsc{CanonParseDiag}_{\Delta}(Q_{\text{str}})$}
    \State \Return \blue{$\{(R_{\alpha,\pi},\kappa,\psi_{\pi}(Q_{\text{str}}))\}$} \Comment{\blue{Malformed SQL never proceeds to $\beta/\gamma$}}
\EndIf

\State \blue{$\mathcal{E} \gets \Sigma \cup \bigcup_{\Phi' \in \Phi.S} \textsc{Export}(\Phi')$  \Comment{Unified environment; Layer~$\gamma$, \S\ref{sec:validation_engine}}}

\Statex \Comment{\textbf{Layer $\beta$: Context-Free Structural Properties}}
\State \blue{$\mathit{AggCtx}$} $\gets (\exists e \in \blue{\textsc{Expressions}(\Pi \cup \Gamma.H)}: \textsc{IsAggregate}(e)) \lor \Gamma.G \neq \emptyset$
\If{\blue{$\mathit{AggCtx}$}}
    \ForAll{$c \in \blue{\textsc{NonAggAttributes}(\Pi \cup \Gamma.H)}$} \Comment{Aggregation scoping}
        \If{$c \notin \Gamma.G$}
            \State \blue{$\kappa \gets \textsc{CanResolve}_{\Delta,A,\mathcal{E}}(c)$}
            \State $V \gets V \cup \{\blue{(R_{\beta,1},\kappa,\psi_{\text{scope}}(c))}\}$
        \EndIf
    \EndFor
\EndIf

\ForAll{$v=(R,x,\psi) \in \textsc{CheckStructural}_{\beta}(\Phi)$}
    \State \blue{$\kappa \gets \textsc{CanResolve}_{\Delta,A,\mathcal{E}}(x)$}
    \State $V \gets V \cup \{\blue{(R,\kappa,\psi)}\}$
\EndFor

\Statex \Comment{\textbf{Layer $\gamma$: Context-Sensitive Semantic Properties}}
\ForAll{$c \in \textsc{Attributes}(\Phi)$} \Comment{Identifier binding}
    \State \blue{$\kappa \gets \textsc{CanResolve}_{\Delta,A,\mathcal{E}}(c)$}
    \If{$\neg\, \textsc{Resolve}(\kappa,\; \Phi.\rho.T,\; \mathcal{E})$}
        \State $V \gets V \cup \{\blue{(R_{\gamma,1},\kappa,}\psi_{\text{bind}}(c,\textsc{SchemaSearch}(c,\mathcal{E}))\blue{)}\}$
    \EndIf
\EndFor

\ForAll{$\text{op}(\mathbf{a}) \in \mathcal{O}(\Phi)$} \Comment{Dialect type checking}
    \State \blue{$\kappa \gets \textsc{CanonSig}_{\Delta,A,\mathcal{E}}(\text{op},\mathbf{a})$}
    \If{$\Theta(\mathbf{a}) \not\sqsubseteq \textsc{Sig}_{\Delta}(\text{op})$}
        \State $V \gets V \cup \{\blue{(R_{\gamma,3},\kappa,}\psi_{\text{type}}(\text{op},\mathbf{a})\blue{)}\}$
    \EndIf
\EndFor

\If{$\neg\, \textsc{IsConnected}(\Phi.\rho.T,\; \Sigma.\mathcal{K}_{\text{fk}})$}
    \State \blue{$\kappa \gets \textsc{CanResolve}_{\Delta,A,\mathcal{E}}(\Phi.\rho.T)$}
    \State \blue{$V \gets V \cup \{{(R_{\gamma,2},\kappa,\psi_{\text{fk}}(\Phi.\rho.T))}\}$}
\EndIf

\ForAll{$\Phi' \in \Phi.S$} \Comment{$P_{\text{recurse}}$: {internal check on parsed nested scopes}}
    \State $V \gets V \cup \textsc{Analyze}(\Phi',\; \blue{\Sigma},\; \Delta)$
\EndFor

\State \Return $V$
\end{algorithmic}
\end{algorithm}

\runinhead{Syntactic Analysis (Layer $\alpha$).} Let $\pi: \text{String} \to \Phi \cup \{\bot_\pi\}$ denote the fault-tolerant parser, where $\bot_\pi$ signals irrecoverable parse failure. \blue{Recoverable token-local diagnostics, such as reserved-keyword escaping, are first normalized by deterministic $\alpha$-operators.} A remaining result $\pi(Q) = \bot_\pi$ constitutes an $\alpha$-violation (e.g., severe token malformation). Only queries satisfying $\pi(Q) \neq \bot_\pi$ proceed to deeper layers.

\runinhead{Context-Free Structural Properties (Layer $\beta$).} This layer verifies the intrinsic logical integrity of $\Phi$, independent of the external $\Sigma$. We formalize Structural Correctness via a predicate $P_{\text{struct}}(\Phi)$ that enforces relational dependencies. \blue{For ORDER BY, the structural layer only enforces output-scope exposure in
dialect/query contexts that require visible ordering keys; ordinary ORDER BY
identifier binding is deferred to the context-sensitive identifier-resolution
rule $R_{\gamma,1}$.}

\runinhead{Context-Sensitive Semantic Properties (Layer $\gamma$).} This layer validates $\Phi$ against a unified environment \blue{$\mathcal{E} = \Sigma \cup \bigcup_{\Phi' \in \Phi.S} \textsc{Export}(\Phi')$} that merges the base schema \blue{$\Sigma$ with each nested scope's exports, where $\textsc{Export}(\Phi')$ exposes a scope $\Phi' \in \Phi.S$ as a named relation (via its alias in $A$) over the output attributes determined by $\Phi'.\Pi$ and $\Phi'.A$, so that $\mathcal{E}$ supplies both the relation names $\mathcal{E}.\mathcal{T}$ and the resolvable attributes (see Algorithm~\ref{alg:analyzer}).}
The contextual correctness predicate $P_{\text{context}}$ verifies four aspects: (i)~every referenced table and column resolves against~$\mathcal{E}$ (identifier binding), (ii)~every operator application is type-safe under the dialect~$\Delta$ (type correctness), (iii)~the queried tables $\Phi.\rho.T$ form a connected component in the foreign key graph (structural integrity), and (iv)~all nested scopes in~$\Phi.S$ are recursively valid. The concrete rules ($R_{\gamma,1}$--$R_{\gamma,3}$) are formalized in Table~\ref{tab:validation-rules}.

\begin{remark}[Complexity of Dialect Configuration $\Delta$]
\label{rem:dialect_cost}
In our formalism, $\Delta$ is a lightweight, stateless
dialect configuration. It defines AST node equivalences (e.g., MySQL's backticks vs.\ standard quotes)\blue{, operator signature bounds, and repair templates}. Transferring \thistool\ to a novel SQL dialect incurs only a constant-size static configuration cost.
Formally, $\Delta$ is not an independent validation layer \blue{in the stratified dependency order $\alpha \prec_{\mathrm{sd}} \beta \prec_{\mathrm{sd}} \gamma$}. \blue{It is a global parameterized context injected into dialect-guarded structural
predicates such as $\chi_{\mathrm{ord}}^{\Delta}(\Phi)$, into the $\gamma$-layer
verification predicates $P_{\mathrm{context}}(\Phi,\Sigma,\Delta)$, and into
the [S-Type] candidate subspace construction $\textsc{Repairs}_{\Delta}$. The same configuration therefore guides both validation and repair.}
\end{remark}

A CQS~$\Phi$ is \blue{valid under the declared rule family} when both structural ($\beta$) and contextual ($\gamma$) predicates hold: $P_{\text{struct}}(\Phi) \land P_{\text{context}}(\Phi, \Sigma, \Delta)$. The \textbf{Violation Set} $V(\Phi, \Sigma, \Delta)$ collects all atomic predicates that fail.

\subsection{Concrete Validation Rules}
\label{sec:validation_rules}

We now specify representative rules that concretize the predicates $P_{\text{struct}}$ and $P_{\text{context}}$. These rules are critical: they define the exact boundary of \blue{bounded runtime invariants}.

\begin{table}[t]
\centering
\footnotesize
\begin{tabular}{@{} lp{12cm}l @{}}
\toprule
\textbf{ID} & \textbf{Rule Definition} & \textbf{Layer} \\
\midrule
$R_{\beta,1}$ &
$\blue{\mathit{AggCtx}(\Phi)\Rightarrow{}} \forall c \in \textsc{\blue{NonAggAttributes}}(\Pi \cup \Gamma.H): c \in \Gamma.G$
\newline
\blue{where $\mathit{AggCtx}(\Phi)\equiv(\exists e\in\textsc{Expressions}(\Pi\cup\Gamma.H):\textsc{IsAggregate}(e))\lor\Gamma.G\neq\emptyset$.}
\newline
(In an aggregation context, every non-aggregate attribute in SELECT or HAVING must appear in GROUP~BY.) & $\beta$ \\

$R_{\beta,2}$ &
\blue{$\chi_{\mathrm{ord}}^{\Delta}(\Phi)
\Rightarrow
\forall c \in \mathrm{Attributes}(\Omega_{\mathrm{ord}}.O):
c \in \mathrm{Attributes}(\Pi) \lor c \in \Gamma.G$}
\newline
\blue{(ORDER BY attributes must be output-visible only in dialect/query contexts
that require visible ordering keys; ordinary ORDER BY identifier binding is
handled by $R_{\gamma,1}$.)}
&
\blue{$\beta/\Delta$} \\

$R_{\beta,3}$ &
\blue{$\forall e \in \mathrm{AggExprs}(\Phi): \neg \mathrm{NestedAggregate}(e)$}
\newline
(Aggregate functions must not be nested, e.g., \texttt{MAX(COUNT(...))} is prohibited.) & $\beta$ \\

\addlinespace[0.3em]

$R_{\gamma,1}$ & \blue{$\forall t \in \rho.T: t.\text{name} \in \mathcal{E}.\mathcal{T}$} 
$\;\land$ 
\blue{$\forall c \in \text{AllAttributes}(\Phi): c.\text{name} \in \textsc{SourceAttributes}(\Phi, \mathcal{E})$} 
\newline (All referenced tables and columns must resolve against the \blue{unified environment~$\mathcal{E}$, i.e., the schema or an enclosing CTE/subquery} scope.) & $\gamma$ \\

$R_{\gamma,2}$ &
$\textsc{IsConnected}(\Phi.\rho.T, \Sigma.\mathcal{K}_{\text{fk}})$
\newline
(Tables in FROM/JOIN must be connected via foreign keys.) & $\gamma$ \\

$R_{\gamma,3}$ &
$\forall \text{op}(\mathbf{a}) \in \mathcal{O}(\Phi): \textsc{Compatible}_{\Delta}(\text{op}, \Theta(\mathbf{a}))$
\newline
(Strict dialect-aware type checking for all operators.) & $\gamma/\Delta$ \\
\bottomrule
\end{tabular}
\caption{Representative hierarchical validation rules. $R_{\beta,1}$ unifies aggregation scoping for both SELECT and HAVING clauses. $R_{\gamma,1}$ unifies table existence and column binding into a single identifier resolution check. $R_{\gamma,3}$ is annotated $\gamma/\Delta$ as its predicates are parameterized by the dialect configuration. \blue{Here $\chi_{\mathrm{ord}}^{\Delta}(\Phi)$ denotes the guard that is true only when the target dialect/query context requires ORDER BY keys to be output-visible.}}
\label{tab:validation-rules}

\end{table}

\runinhead{Validation Rules.}
Table~\ref{tab:validation-rules} formalizes a representative subset of the rules. \blue{We use a single guarded predicate $\chi_{\mathrm{ord}}^{\Delta}(\Phi)$ for ORDER BY exposure: $\chi_{\mathrm{ord}}^{\Delta}(\Phi)$ is true only when the target dialect/query context requires ORDER BY keys to be output-visible, such as DISTINCT or derived-scope export boundaries. The predicate $\chi_{\mathrm{ord}}^{\Delta}(\Phi)$ uses the \code{DISTINCT} setting recorded in $\Omega_{\text{ord}}$.} The structural predicate is defined as the conjunction $P_{\text{struct}}(\Phi) \Leftrightarrow \bigwedge_i R_{\beta,i}(\Phi)$, and the contextual predicate as $P_{\text{context}}(\Phi, \Sigma, \Delta) \Leftrightarrow \bigwedge_j R_{\gamma,j}(\Phi, \Sigma, \Delta)$. This rule-based formalization is inherently extensible: adding support for a new SQL construct or a novel error category requires only defining a new rule~$R_{k}$ and assigning it to the appropriate layer.

\subsection{Verification-Guided Search Space Reduction}
\label{sec:repair_rules}

Unlike traditional deductive program repair~\citep{ProveNFix}, which operates in a deterministic semantic space to synthesize exact patches, our framework addresses the non-deterministic, generative nature of \blue{LLM repair}. The core contribution of \thistool\ is therefore not a strict deductive synthesis engine, but rather a \textbf{Search Space Reduction} mechanism.

When the hierarchical analyzer detects a violation $v \in V$, it does not automatically deduce a single correct fix. Instead, it computes a bounded \textit{Candidate Subspace} $\Omega(v)$ \blue{whose elements are drawn from the schema, dialect, and repair-operator contracts. This narrows the LLM's job from open-ended SQL rewriting to proposing a patch inside an explicit, verifier-checked space.}
The \textbf{Strategic Mapping}~$\mathcal{M}$ formally translates
violations into these constrained subspaces:

\vspace{0.5em}
\noindent\textbf{[S-Bind]} \textit{Identifier Subspace Construction.} When a column~$c$ is unresolvable ($v_{\text{bind}}$), the analyzer restricts the LLM's vocabulary strictly to the set of valid identifiers existing in the target schema context:
\begin{equation}
\Omega_{\text{bind}}(c) = \Big\{ \omega_{\text{replace}}(c, c^*) \;\Big|\; c^* \in \blue{\text{SourceAttributes}}(\Phi, \mathcal{E}) \Big\}
\end{equation}

\noindent\textbf{[S-Type]} \blue{\textit{Dialect Repair Subspace.} For a dialect violation at $\text{op}(\mathbf{a})$, $\mathcal{M}$ instantiates repair templates supplied by $\Delta$:}
\begin{equation}
\Omega_{\text{type}}(\text{op}, \mathbf{a}) = \blue{\textsc{Repairs}_{\Delta}(\text{op},\mathbf{a})}
\end{equation}
\blue{Here, $\textsc{Repairs}_{\Delta}$ returns the finite set of applicable operators instantiated from the configured templates, including type casts and function substitutions. Each operator declares its writable scope and a check that the local edit matches the instantiated template, following Definition~\ref{def:edit-boundary}.}

\noindent\textbf{[S-Join]} \textit{Topology Bridging Subspace.} When the FROM clause tables are topologically disconnected ($v_{\text{integrity}}$), the analyzer traverses the foreign key graph $\Sigma.\mathcal{K}_{\text{fk}}$ to emit legal bridging paths:
\begin{equation}
\Omega_{\text{join}}(T) = \Big\{ \omega_{\text{add\_join}}(t^*, \textsc{FKPred}) \;\Big|\; t^* \in \textsc{Paths}(T, \Sigma.\mathcal{K}_{\text{fk}}) \Big\}
\end{equation}
\blue{When a binding diagnostic identifies a schema table $t$ missing from the query's sources, we apply the same construction to $\Phi.\rho.T \cup \{t\}$.}

\noindent \blue{\textbf{[S-Group/S-Order] Structural Scoping Subspaces.}
The two clause-scoping $\beta$ repairs induce singleton subspaces:
$R_{\beta,1}$ yields
$\Omega_{\mathrm{group}}(v_{\mathrm{scope}}(c))=\{\omega_{\mathrm{add\_group}}(c)\}$
for each $c\in \mathrm{NonAggAttributes}(\Pi\cup\Gamma.H)\setminus\Gamma.G$, and
$R_{\beta,2}$ (under $\chi_{\mathrm{ord}}^{\Delta}$) yields
$\Omega_{\mathrm{order}}(v_{\mathrm{order}}(c))=\{\omega_{\mathrm{expose\_order}}(c)\}$.
Their edit boundaries are declared in Table~\ref{tab:operator-edit-boundaries}; all non-target CQS components are frame-preserved.
}

\noindent\blue{\textbf{[S-LiftAgg] Nested-Aggregate Lifting Subspace.}
For $v_{\mathrm{nested}}(e)$ from $R_{\beta,3}$, decompose
$e=A_{\mathrm{out}}(C[A_{\mathrm{in}}(x)])$. Let
$K_e=\Gamma.G\cup\mathrm{FreeAttrs}(C)$ and let $a_e$ be a fresh alias for
$A_{\mathrm{in}}(x)$. Then
\begin{equation}
\Omega_{\mathrm{lift}}(e)
=
\left\{
\omega_{\mathrm{lift\_agg}}(e,m,(K_e,a_e,s_e))
\;\middle|\;
\begin{array}{l}
m\in\{\mathrm{CTE},\mathrm{Subquery}\},\\
s_e \text{ computes } a_e:=A_{\mathrm{in}}(x) \text{ under key } K_e,\\
\mathrm{RelSupp}(e,\Phi)\preceq_{\mathrm{prov}}\mathrm{RelSupp}(s_e)
\end{array}
\right\}.
\end{equation}
}

These bounding equations formalize our neuro-symbolic integration: the \blue{candidate operators} are rigorously defined by the static environment ($\Sigma, \Delta$), drastically pruning the search space. The LLM's role is thus confined to \blue{proposing SQL consistent with these candidate operators; the monitor decides whether the resulting CQS transition is accepted.}

\begin{remark}[Completeness and Extensibility Bounds]
\label{rem:operator_bounds}
The operations $\omega \in \Omega$ are \blue{bounded by the declared} $\alpha,\beta,\gamma$ property violations \blue{and the dialect configuration}. \blue{Adding a new repair class requires three local extensions: a validation rule, a candidate-subspace constructor, and an operator contract describing its allowed CQS edits. These rules are declarative YAML configurations, allowing quick extensibility to new error classes and dialect idiosyncrasies without altering the monitor itself.}
\end{remark}

Table~\ref{tab:repair_by_category} provides the concrete instantiation of this search space reduction across the three error categories, showing the full pipeline from error detection (What?), to bounding evidence (Why?), to the constrained prompt specification delivered to the LLM (How?).

\begin{table}[t]
\centering
\footnotesize
\scalebox{0.85}{
\begin{tabular}{@{}p{0.35\textwidth}p{0.35\textwidth}p{0.4\textwidth}@{}}
\toprule
\textbf{Error Taxonomy \& Example (What?)} & \textbf{Evidence \& Repair Reasoning (Why?)} & \textbf{Repair Specification (How?)} \\
\midrule
\textbf{Syntactic} 

Condition: $\alpha: \pi(Q) = \bot_\pi$ 

\textbf{Example:} \texttt{SELECT order\_id FROM order} 

\textbf{Error:} near ``order'': syntax error
& 
\textbf{Grammar-Driven Hypothesis}: Use fault-tolerant parsing to identify token-level errors (spelling, escaping). Apply string similarity (Levenshtein distance) to find the intended SQL keyword from grammar rules. &
Apply direct token replacement based on fault-tolerant parsing and string similarity.

\textbf{Example:} \textit{Replace "order" with `order' based on SQLite grammar matching} 
\\[0.5em]
\midrule
\textbf{Context-Free Structural} 

Condition: $\beta: \neg\, P_{\text{struct}}(\Phi)$ 

\textbf{Example: }\texttt{SELECT \blue{DISTINCT} account\_id FROM account ORDER BY order\_id ...}

\textbf{Error:} no such column: order\_id
& 
\textbf{Intra-Query Structural Analysis}: Detect violations of SQL's built-in semantic rules (e.g., aggregation scope, alias conflicts, cross-clause references) by analyzing query structure. Generate specific prompts telling LLM which rule is violated and how to fix. &
\textbf{Prompts:} Generate structured prompts detailing the violation and required action.

\textbf{Example:} \blue{\textit{``ORDER-BY EXPOSURE VIOLATION: The target dialect requires an output-visible ORDER BY key under DISTINCT. Use an existing projected expression or alias as the ordering key. Keep the projection unchanged.''}} \\[0.5em]

\midrule
\textbf{Context-Sensitive Semantic} 

Condition: $\gamma: \neg\, P_{\text{context}}(\Phi, \Sigma, \Delta)$ 

\textbf{Example:} \texttt{SELECT Patient\_ID FROM Laboratory} 

\textbf{Error:} no such column: Laboratory.Patient ID
& 
\textbf{Schema-Grounded Entity Search}: Search database schema $\Sigma$ for valid alternatives to non-existent tables/columns. Use foreign key relationships and string similarity as evidence to provide LLM with ranked replacement suggestions. &
\textbf{Prompts:} Generate schema-grounded prompts with ranked replacement suggestions.

\textbf{Example:} \textit{``COLUMN NOT FOUND: Column `Patient\_ID' does not exist in table `Laboratory'. Hint: REPLACE `Laboratory.Patient\_ID' WITH `Laboratory.ID' based on schema''} \\[0.5em]
\bottomrule
\end{tabular}
}
\caption{Repair Paradigms by Formal Error Category}
\label{tab:repair_by_category}

\end{table}

\section{Operational Semantics of Property-Guided Repair}
\label{sec:repair_formalism}

Given an invalid CQS with $V \neq \emptyset$, our framework transforms the unstructured string manipulation of EF into a state-transition system. \blue{A verify-before-apply temporal monitor governs each LLM-generated patch: it re-canonicalizes the candidate, re-runs the analyzer, and accepts the transition only if the violation count strictly decreases, no previously resolved violation recurs, and an admissible operator plan satisfies the edit-boundary and preservation checks.}

\subsection{The Repair Loop}
\label{sec:repair_loop}

The repair process proceeds iteratively over CQS instances. 
\blue{Algorithm~\ref{alg:analyzer} presents the complete analysis from a raw SQL string, including parsing and failure reporting. For an already parsed CQS, $\textsc{Analyze}(\Phi,\Sigma,\Delta)$ denotes its $\beta/\gamma$ validation portion, starting with construction of the environment~$\mathcal{E}$. Repair-loop and nested-scope calls use this portion directly. We omit the iteration index $i$ when describing a single repair transition.} The system is defined as follows:
\begin{itemize}
    \item $\textsc{Analyze}(\Phi, \Sigma, \Delta) \to \mathcal{P}(E)$, the hierarchical static analyzer (Algorithm~\ref{alg:analyzer}), where $E$ is the space of detectable property violations.
    \item $\mathcal{M}: E \to \mathcal{P}(\Omega)$, the strategic mapping from violations to candidate subspaces (\S~\ref{sec:repair_rules}).
\end{itemize}

\begin{definition}[\blue{Canonical Violation Identity}]
\label{def:violation-identity}
\blue{The analyzer emits each violation as a tuple:
\begin{equation} 
{\label{eq:violation-tuple}
v=(R,\kappa,\psi),
}\end{equation}
where $R \in \{R_{\alpha,i},R_{\beta,j},R_{\gamma,k}\}$} identifies the failed rule (Table~\ref{tab:validation-rules}), \blue{$\kappa$ is the canonical identity used for equality testing across repair iterations, and $\psi$ is the diagnostic payload used for prompting. For $\alpha$-layer parse violations, $\kappa$ is derived from the canonical parser diagnostic because no resolved identifier or expression may exist. For $\beta/\gamma$ violations, $\kappa$ is computed over CQS objects after dialect normalization, alias resolution, and CTE/subquery export resolution; Algorithm~\ref{alg:analyzer} implements this normalization with $\textsc{CanResolve}_{\Delta,A,\mathcal{E}}$. For identifier-binding violations, $\kappa$ records the resolved scope, relation, and attribute identity; for type violations, it records the canonical operator signature; for $\beta$-layer structural violations, it records the canonical clause/expression identity after alias expansion. Two violations are equal iff both $R$ and $\kappa$ match. This prevents false recurrence or false progress caused only by alias renaming, CTE rewriting, or dialect quoting.}
\end{definition}

\blue{Let $\mathcal{H}_i$ denote the resolved-violation history available before iteration~$i$, a set of resolved violations compared by their canonical identity (Definition~\ref{def:violation-identity}). It is initialized as $\mathcal{H}_0=\emptyset$ and updated after an accepted transition $\Phi_i\to\Phi_{i+1}$ by}
\begin{equation}
{\label{eq:H-update}
\blue{\mathcal{H}_{i+1}=\mathcal{H}_i\cup (V_i\setminus V_{i+1}),}
}\end{equation}
\blue{where $V_i=\textsc{Analyze}(\Phi_i,\Sigma,\Delta)$. Thus $\mathcal{H}_i$ contains the violations that were present in an earlier accepted state and have since been removed. When the index is omitted, $\mathcal{H}$ denotes the current history.}

At each iteration, the engine \blue{analyzes the current CQS~$\Phi$ to obtain its violation set $V=\textsc{Analyze}(\Phi,\Sigma,\Delta)$ and derives the bounded candidate subspace $\Omega_{\text{cand}}=\bigcup_{v\in V}\mathcal{M}(v)$ induced by these diagnostics. Treating the LLM as an untrusted repair proposer, the engine compiles the current violations, the candidate operators with their declared edit boundaries, and the resolved-violation history~$\mathcal{H}$ into a constrained specification, from which the proposer returns a candidate CQS~$\Phi_{\text{cand}}$. The monitor accepts the transition only when the four runtime checks below hold, consolidated into the three conjuncts of the acceptance condition~\eqref{eq:acceptance}; otherwise it rejects the candidate, leaves~$\Phi$ unchanged, and re-prompts the proposer within a finite retry budget}:

\begin{enumerate}[leftmargin=*, nosep]
    \item \textbf{Monotonic Progress.}  
    The candidate violation count must strictly decrease: \blue{$|V_{\text{cand}}| < |V|$}.
    \item \textbf{Regression Safety.} \blue{No previously resolved violation may recur: $V_{\text{cand}} \cap \mathcal{H} = \emptyset$, where $\mathcal{H}$ is the history of Eq.~\eqref{eq:H-update}; because it accumulates over the whole trajectory rather than relative to the initial set, it also blocks reintroducing a violation that was resolved mid-repair}.
    \item \textbf{\blue{Edit-Boundary Check.}} 
    \blue{The observed CQS delta may touch only fields authorized by some operator plan~$\Lambda\subseteq\Omega_{\text{cand}}$.}
    \item \textbf{\blue{Frame Preservation.}} \blue{Every CQS component outside those authorized fields must be preserved, except for explicitly checked rewrites such as CTE lifting that preserve tables and filters by subsumption.}
\end{enumerate}

\begin{definition}[\blue{Operator Edit Boundary}]
\label{def:edit-boundary}
Let \blue{$\mathcal{X} = \{\Pi, \rho.T, \rho.J, \rho.F, \Gamma.G, \Gamma.H,  \Omega_{\text{ord}}, A, S\}$} be the set of protected CQS components \blue{monitored by the repair loop (written $\mathcal{X}$ to avoid clashing with the schema columns $\mathcal{C}$ of $\Sigma$). Each instantiated repair operator~$\omega\in\Omega$ declares the rule it targets, $\mathsf{Rule}(\omega)$, the CQS components it may modify, $\textsc{Scope}(\omega)\subseteq\mathcal{X}$, and an executable local predicate $\textsc{Check}(\omega,\Phi,\Phi')$. The local predicate captures operator-specific applicability and preservation obligations, such as projection preservation for \code{ORDER BY} repair or relational-support preservation for CTE lifting. For node-local repairs, $\textsc{Scope}(\omega)$ contains the CQS components enclosing the targets fixed at instantiation, while $\textsc{Check}$ constrains the local rewrites at those targets. At iteration~$i$, the current violation set~$V_i$ determines the candidate operator subspace~$\Omega_i$; a repair plan $\Lambda_i\subseteq\Omega_i$ records the operators that can authorize the observed transition. The monitor checks:}

\begin{equation}
\label{eq:invariantok}
\textsc{InvariantOK}(\Phi,\Phi',\blue{\Lambda_i})
=
\left(
\blue{
\bigwedge_{\omega\in\Lambda_i}
 \textsc{Check}(\omega,\Phi,\Phi')}
\right)
\land
\left(
\bigwedge_{X\in\blue{\mathcal{X}}\setminus \blue{\bigcup_{\omega\in\Lambda_i} \textsc{Scope}(\omega)}}
X(\Phi) \blue{\preceq_X} X(\Phi')
\right),
\end{equation}
\blue{where the first conjunct enforces the local predicate declared by each selected operator, and the second conjunct preserves every CQS component outside the selected operators' scopes. Here $\preceq_X$ is equality for ordinary fields and provenance-preserving subsumption for scoped rewrites such as CTE lifting. The witness plan~$\Lambda_i$ ranges over subsets of the current candidate subspace~$\Omega_i$: a transition is accepted only if some admissible plan drawn from~$\Omega_i$ explains the observed edit, so any mutation that no declared operator scope can account for is rejected.}
\end{definition}

\blue{Concretely, for the representative $\beta$-layer rules in Table~\ref{tab:validation-rules}, edit boundaries are declared directly from the structural rule that fired. For $R_{\beta,1}$, $\omega_{\mathrm{add\_group}}(c)$ applies only in an aggregation context and has scope $\{\Gamma.G\}$. For $R_{\beta,2}$, $\omega_{\mathrm{expose\_order}}(c)$ has scope over $\Omega_{\text{ord}}$ and aliases. It rewrites $\Omega_{\text{ord}}.O$ to use expressions in $\Pi$ or their aliases in $A$, while preserving $\Pi$ and the DISTINCT modifier. For $R_{\beta,3}$, $\omega_{\mathrm{lift\_agg}}(e)$ may scope over $\Pi$, $\rho.T$, $\Gamma.G$, $\Gamma.H$, and $S$ to lower a nested aggregate into a CTE or subquery; the outer $\rho.J$ and $\rho.F$ stay frame-preserved, and any $\rho.T$ rewrite must go through a derived scope that preserves the original relational support needed by the aggregate and must pass full re-analysis.}

\begin{algorithm}[tbp]
\footnotesize
\caption{Violation-Driven Repair Plan Synthesis \blue{with a Verify-Before-Apply Temporal Monitor}}
\label{alg:synthesis}
\begin{algorithmic}[1]
\Require Initial CQS $\Phi$, Schema $\Sigma$, Dialect $\Delta$, \blue{Retry budget $k_{\max}$}
\Ensure \blue{Current} CQS $\Phi$
\State $V \gets \textsc{Analyze}(\Phi, \Sigma, \Delta)$ \Comment{Algorithm~\ref{alg:analyzer}}
\If{$V = \emptyset$}
    \State \Return $\Phi$ \Comment{Already valid}
\EndIf
\State \blue{$\mathcal{H} \gets \emptyset$} \Comment{\blue{Resolved-violation history; never reintroduced}}

\While{$V \neq \emptyset$}
\Statex \Comment{\textbf{Phase 1: Symbolic Search Space Reduction}}
    \State $\Omega_{\text{cand}} \gets \blue{\bigcup_{v\in V}\mathcal{M}(v)}$ \Comment{\blue{Bounded candidate subspace; its operators bound the admissible edits}}
    \State \blue{$\textit{rej} \gets \emptyset$}

    \For{$j = 1$ \textbf{to} $k_{\max}$}
\Statex \Comment{\textbf{Phase 2: Constrained Neural Execution}}
        \State $\Phi_{\text{cand}} \gets \textsc{LLM.Generate}(\textsc{BuildSpec}(\Phi, V, \Omega_{\text{cand}}\blue{, \mathcal{H}, \textit{rej}}))$ \Comment{Neural repair proposer}

\Statex \Comment{\textbf{\blue{Phase 3 : Verify-Before-Apply Temporal Monitor}}}
        \State $V_{\text{cand}} \gets \textsc{Analyze}(\Phi_{\text{cand}}, \Sigma, \Delta)$
        \State $\textit{progress} \gets |V_{\text{cand}}| < |V|$ \Comment{Monotonic progress}
        \State $\textit{no\_regress} \gets \blue{V_{\text{cand}} \cap \mathcal{H}} = \emptyset$ \Comment{Regression safety}
        \State $\textit{preserved} \gets \blue{\exists \Lambda\subseteq\Omega_{\text{cand}}:\; \textsc{InvariantOK}(\Phi, \Phi_{\text{cand}}, \blue{\Lambda})}$ \Comment{Invariant preservation}
        \If{$\textit{progress} \land \textit{no\_regress} \land \textit{preserved}$}
            \State \blue{$\mathcal{H} \gets \mathcal{H} \cup (V \setminus V_{\text{cand}})$} \Comment{\blue{Record newly resolved violations}}
            \State \blue{$(\Phi, V) \gets (\Phi_{\text{cand}}, V_{\text{cand}})$} \Comment{\blue{Accepted transition updates state}}
            \State \blue{\textbf{break}}
        \Else
            \State \blue{$\textit{rej} \gets$ diagnostics of failed monitor checks}
        \EndIf
    \EndFor

    \If{no candidate was accepted}
        \State \Return \blue{$\Phi$} \Comment{\blue{Retry budget exhausted; keep current CQS}}
    \EndIf
\EndWhile

\State \Return \blue{$\Phi$}

\end{algorithmic}

\end{algorithm}

A transition is accepted only when \blue{monitored checks hold;} otherwise, the candidate is rejected and the LLM is re-prompted with refined diagnostics. \blue{Here, $\textit{rej}$ stores the diagnostics of failed monitor checks for the next \textsc{BuildSpec} call.}
Algorithm~\ref{alg:synthesis} proceeds in three phases and \blue{ maintains the history~$\mathcal{H}$}: \textbf{Phase~1 (Symbolic Search Space Reduction):} The engine runs $V=\textsc{Analyze}(\Phi,\Sigma,\Delta)$ and instantiates $\mathcal{M}(v)$ for every violation $v \in V$, producing the bounded candidate subspace~$\Omega_{\text{cand}}$ \blue{and the edit boundaries declared by its operators}. \textbf{Phase~2 (Constrained Neural Execution):} $\textsc{BuildSpec}$ compiles the current CQS, violation diagnostics, candidate subspace, operator edit boundaries, $\mathcal{H}$\blue{, and $\textit{rej}$} into a constrained specification. The LLM ($\textsc{LLM.Generate}$) proposes a candidate CQS $\Phi_{\text{cand}}$. \textbf{Phase~3 (Verify-Before-Apply):} The \blue{Verify-Before-Apply Temporal Monitor} re-validates $\Phi_{\text{cand}}$ via the static analyzer and accepts the transition only if the acceptance condition~\eqref{eq:acceptance} holds\blue{, including the existence of an admissible operator plan witnessing the edit boundary; on acceptance, the newly resolved violations are added to the resolved-violation history~$\mathcal{H}$.}

\begin{table*}[t]
\centering
\footnotesize
\renewcommand{\arraystretch}{1.25}
\begin{tabular}{@{}l p{0.12\textwidth} p{0.25\textwidth} p{0.47\textwidth}@{}}
\toprule
\textbf{Rule} & \textbf{Operator $\omega$} & \textbf{$\textsc{Scope}(\omega)$} & \textbf{$\textsc{Check}(\omega,\Phi,\Phi')$} \\
\midrule
$R_{\beta,1}$ & $\omega_{\mathrm{add\_group}}$ & $\{\Gamma.G\}$ & adds $c$ to $\Gamma.G$; other components preserved. \\
$R_{\beta,2}$ & $\omega_{\mathrm{expose\_order}}$ & $\{\Omega_{\text{ord}},A\}$ & \blue{rewrites $\Omega_{\text{ord}}.O$ to use expressions in $\Pi$ or their aliases in $A$, while preserving $\Pi$ and the DISTINCT modifier.} \\
$R_{\beta,3}$ &
$\omega_{\mathrm{lift\_agg}}$ &
$\{\Pi, \rho.T, \Gamma.G, \Gamma.H,S\}$ &
\blue{lifts the inner aggregate into a CTE/subquery as specified
in [S-LiftAgg], then uses its exported result in the original
outer expression. Preserves the original joins and filters,
including those moved into the derived scope, and the visible
output schema. Requires full re-analysis.} \\
\addlinespace[0.35em]
$R_{\gamma,1}$ & $\omega_{\text{replace}}$ & \blue{components containing target references $c$} & \blue{rebinds targeted references $c\!\mapsto\!c^*$, with $c^*\!\in\!\textsc{SourceAttributes}(\Phi,\mathcal{E})$; query structure unchanged.} \\
$R_{\gamma,2}$ & $\omega_{\text{add\_join}}$ & $\{\rho.T,\rho.J\}$ & adds an FK-path bridge; $\Pi,\Gamma,\Omega_{\text{ord}}$ preserved. \\
$R_{\gamma,3}$ & $\omega_{\text{cast}}$ & \blue{components containing the targeted operator node} & \blue{wraps the offending operand at the target operator in a cast to a target type $\tau^*$ allowed by $\Delta$; operand expressions preserved.} \\
\bottomrule
\end{tabular}
\caption{\blue{Edit boundary contracts for the repair operators. Each operator declares its writable scope $\textsc{Scope}(\omega)$ and check predicate $\textsc{Check}(\omega,\Phi,\Phi')$ (Def.~\ref{def:edit-boundary}).}}
\label{tab:operator-edit-boundaries}

\end{table*}

\subsection{Progress and Preservation}
\label{sec:soundness}

\blue{The monitor enforces }\textbf{bounded soundness} \blue{over accepted transitions.}
We consolidate the \blue{checks from \S\ref{sec:repair_loop}} into a single acceptance condition. A candidate transition from $\Phi_i$ to $\Phi_{i+1}$ is accepted if:
\begin{equation}
\label{eq:acceptance}
|V_{i+1}| < |V_i| \;\land\;
V_{i+1}\blue{\cap \mathcal{H}_i} = \emptyset \;\land\;
\blue{\exists \Lambda_i\subseteq\Omega_i:\;}
\textsc{InvariantOK}(\Phi_i,\Phi_{i+1},\blue{\Lambda_i})\blue{.}
\end{equation}

\begin{definition}[\blue{Bounded Soundness}]
\blue{Given a rule family $\mathcal{R}$ and the operator contracts of Definition~\ref{def:edit-boundary}, a repair trace is \emph{bounded-sound} if every accepted transition $\Phi_i \rightarrow \Phi_{i+1}$, with $V_i=\textsc{Analyze}_{\mathcal{R}}(\Phi_i,\Sigma,\Delta)$, satisfies}
\begin{equation}
\label{eq:bounded-soundness}
\blue{
|V_{i+1}| < |V_i|
\;\land\;
V_{i+1} \cap \mathcal{H}_i = \emptyset
\;\land\;
\exists \Lambda_i\subseteq\Omega_i:\;
\textsc{InvariantOK}(\Phi_i,\Phi_{i+1},\Lambda_i).
}
\end{equation}
\blue{This property is bounded to violations detectable by $\mathcal{R}$ and does not imply semantic equivalence to the gold SQL or preservation of user intent.}
\end{definition}

\begin{invariant}[Monotonic Progress and Preservation]
\label{thm:repair-soundness-main}
Let $\Phi_0$ be an initial invalid CQS. Every accepted repair transition in the sequence $\Phi_0,\Phi_1,\dots$ satisfies the acceptance condition~\eqref{eq:acceptance}. Consequently, \blue{accepted transitions strictly decrease the violation count, reintroduce no previously resolved violation, respect an edit boundary witnessed by operators in the current candidate subspace, and preserve every non-target CQS component.}
\end{invariant}

\begin{proof}
\textit{Progress and Regression Safety:}
By construction of $\mathcal{M}$, each candidate subspace $\Omega(v)$ enumerates \blue{edits} drawn from the validated schema $\Sigma$ and dialect $\Delta$ (e.g., $c^*\in\textsc{SourceAttributes}(\Phi,\mathcal{E})$ for [S-Bind]). However, the LLM may deviate from the \blue{constrained specification. The monitor therefore} re-diagnoses the \textit{entire} CQS via Algorithm~\ref{alg:analyzer} and accepts only if \blue{$|V_{i+1}|<|V_i|$ and $V_{i+1}\cap \mathcal{H}_i=\emptyset$. Since $\mathcal{H}_i$ records every violation removed by an earlier accepted transition, this rejects candidates that fix one violation while reintroducing a resolved one, even if that violation was absent from the immediately preceding state.}

\textit{\blue{Edit and Frame Preservation:}}
\blue{Algorithm~\ref{alg:synthesis} accepts the candidate only if there exists $\Lambda_i\subseteq\Omega_i$ such that $\textsc{InvariantOK}(\Phi_i,\Phi_{i+1},\Lambda_i)$ holds. Its operator-check conjunct rejects violations of the witness operators' local obligations, and its frame conjunct rejects mutations to non-target CQS fields.}
\end{proof}

The proof of \blue{Runtime Invariant}~\ref{thm:repair-soundness-main} relies on two complementary mechanisms: the design-time construction of $\mathcal{M}$, which \blue{narrows what the LLM should produce}, and the \blue{temporal} monitor in Algorithm~\ref{alg:synthesis}, which \blue{enforces what the system actually accepts.}

\subsection{\blue{Bounded Halting}}

By enforcing 
\blue{a strictly decreasing violation count, the monitor bounds the number of accepted transitions, while the finite retry budget $k_{\max}$ bounds the rejected proposals tried at each state; once a state exhausts its budget, the algorithm returns the current CQS without overwriting it. We make this precise with} the potential function \blue{$\mu(\Phi) = |V(\Phi, \Sigma, \Delta)| \in \mathbb{N}$.}

\blue{
\begin{definition}[Bounded Halting]
\label{def:bounded-halting}
A repair loop has bounded halting under retry budget $k_{\max}$ if, for any initial CQS and any LLM behavior, it terminates after finitely many accepted transitions and finitely many LLM invocations, returning either a zero-violation CQS or the last accepted one. This is a halting property; it does not guarantee successful repair.
\end{definition}
}

\begin{invariant}[Bounded Halting of the Repair Loop]
\label{inv:bounded-halting}
 For any initial \blue{CQS $\Phi_0$ and any finite $k_{\max}$, Algorithm~\ref{alg:synthesis} performs at most $\mu(\Phi_0)$ accepted transitions and at most $\mu(\Phi_0)\cdot k_{\max}$ LLM invocations.}
\end{invariant}

\begin{proof}
\blue{By Runtime Invariant~\ref{thm:repair-soundness-main}, each accepted transition strictly decreases the nonnegative integer $\mu$, allowing at most $\mu(\Phi_0)$ accepted transitions. Only states with $\mu>0$ invoke the LLM, each at most $k_{\max}$ times, giving at most $\mu(\Phi_0)\cdot k_{\max}$ invocations.}
\end{proof}


\subsection{\detalign\ for Empty Results}
\label{sec:delta_layer}
While the preceding layers form a static analysis hierarchy, some failure modes require \textit{data-level evidence}. We explicitly isolate \detalign\ as a distinct post-validation step to preserve the static purity of the core analysis.

\begin{algorithm}[!htbp]
\footnotesize
\caption{\detalign\ for Empty Results}
\label{alg:refinement}
\begin{algorithmic}[1]
\Require Valid CQS $\Phi$ with $V(\Phi,\Sigma,\Delta)=\emptyset$, Database $\mathcal{D}$, \blue{Schema $\Sigma$, Dialect $\Delta$}
\Ensure Aligned CQS $\Phi'$ or $\bot_{\text{nc}}$

\Statex \Comment{\textbf{Step 1: Identify failing filter predicates}}
\State $F_{\text{fail}} \gets \{ f \in \Phi.\rho.F \mid |\textsc{Eval}(f,\mathcal{D})| = 0 \}$

\Statex \Comment{\textbf{Step 2: Classify by correctability}}
\State $F_{\text{col}} \gets \{ f=(c,\circ,v) \in F_{\text{fail}} \mid \blue{\textsc{IsColumnRef}(c)\land \textsc{IsLiteral}(v)} \}$ \Comment{Column mismatch}
\State $F_{\text{val}} \gets F_{\text{fail}} \setminus F_{\text{col}}$ \Comment{Value boundary: uncorrectable}

\If{$F_{\text{val}} \neq \emptyset$}
    \State \Return $\bot_{\text{nc}}$ \Comment{Conservative abstention}
\EndIf

\Statex \Comment{\textbf{Step 3: Deterministic column alignment}}
\State \blue{$\mathcal{E} \gets \Sigma \cup \bigcup_{s\in\Phi.S}\textsc{Export}(s)$} \Comment{\blue{Scope-visible environment}}
\State $\Phi' \gets \Phi$
\ForAll{$f = (c,\circ,v) \in F_{\text{col}}$}
    \State $\mathcal{C}_{\text{cand}} \gets \{ c^* \in \blue{\textsc{SourceAttributes}(\Phi',\mathcal{E}) \setminus \{c\}} \mid \blue{\textsc{Compatible}(\textsc{Type}_{\mathcal{E}}(c^*),\circ,\textsc{Type}(v)) \land} |$
    \Statex \hspace{\algorithmicindent}$\textsc{Eval}((c^*,\circ,v),\mathcal{D})| > 0 \blue{\land \textsc{Analyze}(\Phi'[\rho.F:f\mapsto(c^*,\circ,v)],\Sigma,\Delta)=\emptyset} \}$
    \If{$\mathcal{C}_{\text{cand}} = \emptyset$}
        \State \Return \blue{$\bot_{\text{nc}}$} \Comment{No data-supported column candidate}
    \EndIf
    \State \blue{$c' \gets \textsc{First}_{\preceq_{\text{scope}}}(\mathcal{C}_{\text{cand}})$} \Comment{\blue{Canonical scope-order selection}}
    \State \blue{$\Phi' \gets \Phi'[\rho.F:f\mapsto(c',\circ,v)]$} \Comment{Substitute attribute}
 \EndFor

\If{$\textsc{Analyze}(\Phi',\Sigma,\Delta) \neq \emptyset \blue{\lor |\textsc{Eval}(\Phi',\mathcal{D})| = 0}$}
    \State \Return \blue{$\bot_{\text{nc}}$} \Comment{Reject invalid or still-empty alignment}
\EndIf
\State \Return $\Phi'$
\end{algorithmic}
\end{algorithm}

When a query is \blue{valid under the declared static rule family} ($V=\emptyset$) but returns an empty result, \blue{the failure is outside the static rule family}. For such cases (Figure~\ref{fig:archi_design}c), we apply \detalign, \blue{a conservative post-validation heuristic using data-level evidence.}
If a filter predicate \blue{$f=(c,\circ,v)$} yields no tuples, the \textbf{Alignment Operator} $\omega_{\text{align}}(c)$ \blue{constructs attribute-substitution candidates $c^*$ while keeping $\circ$ and $v$ fixed. A candidate is admissible only if $c^*\in\textsc{SourceAttributes}(\Phi,\mathcal{E})$, $|\textsc{Eval}((c^*,\circ,v),\mathcal{D})|>0$, and $\textsc{Analyze}(\Phi[\rho.F:f\mapsto(c^*,\circ,v)],\Sigma,\Delta)=\emptyset$. Algorithm~\ref{alg:refinement} then selects the first admissible candidate in canonical scope order, or returns $\bot_{\text{nc}}$ if none exists. Thus \detalign\ is reproducible and regression-blocking, but does not certify agreement with the user's unstated intent.}

\subsection{Case Studies}
\label{sec:case_studies}
To concretely illustrate the operation of \thistool\, we examine a real-world, highly complex context-sensitive failure from the BIRD SQLite Standard and PostgreSQL Mini-dev benchmark.

\subsubsection{Case 1:}
Consider the following intermediate LLM generation that structurally parses (passing $\alpha$) but harbors subtle PostgreSQL-specific dialect incompatibilities and identifier binding errors:

\begin{lstlisting}[language=SQL, basicstyle=\ttfamily\small]
SELECT ROUND(
    (CAST(COUNT(DISTINCT CASE WHEN y.consumption > 46.73 
        THEN y.customerid END) AS FLOAT) / 
     CAST(COUNT(DISTINCT y.customerid) AS FLOAT)) * 100, 2)
FROM yearmonth y JOIN customers c ON y.customerid = c.customerid;
\end{lstlisting}
Under standard execution feedback, this query immediately crashes the PostgreSQL engine, emitting a dense type-casting stack trace (e.g., \texttt{function round(double precision, integer) does not exist}). The error message conflates the type incompatibility with the function signature, providing no indication that a \textit{separate} identifier binding error also exists. The LLM, prompted only with this opaque trace, must guess both the root cause and whether additional errors are lurking---a combinatorial guessing game that frequently leads to partial fixes and regressions.
In Algorithm~\ref{alg:analyzer}, the query is canonicalized into~$\Phi$ and validated layer-by-layer. The $\alpha$ and $\beta$-layers pass. The $\gamma$-layer semantic analysis discovers two \textit{orthogonal} violations that EF conflates into a single crash:

\begin{enumerate}
    \item \textbf{Identifier Binding ($v_{\text{bind}}$)}: $\neg \exists t \in T \text{ s.t. } (t, \text{customerid}) \in \Sigma.\mathcal{C}$. The schema requires precise case sensitivity (\code{CustomerID}). A schema-wide search locates the correct identifier and emits a targeted replacement suggestion.
    \item \textbf{Dialect Conformance ($v_{\text{type}}$)}: The dialect parameter $\Delta_{\text{Postgres}}$ strictly forbids the \texttt{ROUND()} function from accepting an \texttt{AS FLOAT} double-precision cast, requiring an exact \texttt{AS NUMERIC} parameter signature. This is detected purely from the dialect configuration, without executing the query.
\end{enumerate}

\runinhead{\blue{Synthesis with the Temporal Monitor.}}
The Symbolic Engine isolates both violations $v_1, v_2 \in V$, bypasses the database entirely, and transitions to Algorithm~\ref{alg:synthesis}. For $v_1$, it constructs the candidate subspace $\Omega_{\text{bind}} = \{ \omega_{\text{replace}}(\text{customerid}, \text{CustomerID}) \}$; for $v_2$, the type-cast subspace $\blue{\Omega_{\text{type}}} = \{ \omega_{\text{cast}}(\text{FLOAT}, \text{NUMERIC}) \}$. The composite specification is passed to the LLM, which applies the exact AST patches. The temporal monitor confirms \blue{that the canonical violation set strictly shrinks to } $V(\Phi'', \Sigma, \Delta) = \emptyset$ and \blue{that non-target CQS fields are preserved}. In exactly \textbf{one \blue{accepted} transition} and with \textbf{zero database executions}, the query \blue{reaches zero statically detectable} violations. An EF pipeline would require at least two round-trips (one per error discovered sequentially), with no \blue{monitor-checked boundary} against regressions between rounds.

\subsubsection{Case 2:}
Consider a SQLite query generated for the question \textit{``What is the annual average number of sets released between 2012--2015, and the most common card language?''}:
\begin{lstlisting}[language=SQL, basicstyle=\ttfamily\small]
SELECT AVG(sets.id) AS average_sets,
       MAX(COUNT(set_translations.language)) AS common_language
FROM sets JOIN set_translations
    ON sets.code = set_translations.setCode
WHERE sets.releaseDate BETWEEN '2012-01-01' AND '2015-12-31'
GROUP BY set_translations.language
ORDER BY COUNT(set_translations.language) DESC LIMIT 1;
\end{lstlisting}
The $\alpha$-layer passes (no syntax errors). The $\beta$-layer detects a \textbf{structural scoping violation} ($R_{\beta,3}$): the expression \texttt{MAX(COUNT(...))} nests one aggregate inside another, which is prohibited in standard SQL. This is a context-free structural error ($v_1 = v_{\text{scope}}$, layer~$\beta$) that requires restructuring the query into subqueries or CTEs.
Under EF, the engine reports only \texttt{misuse of aggregate function COUNT()}---an opaque message that gives no indication of the required restructuring strategy. The LLM must independently infer that the fix involves CTE decomposition, a non-trivial architectural transformation. Moreover, EF cannot foresee the dialect issue that will surface \textit{after} the restructuring; it discovers errors one at a time through sequential executions.

\runinhead{\blue{Edit Boundary Checks.}}
\blue{Because $R_{\beta,3}$ requires aggregate lifting, the monitor allows $\omega_{\mathrm{lift\_agg}}$ to edit the projection, data-source, aggregation, and nested-scope fields declared in $\textsc{Scope}(\omega_{\mathrm{lift\_agg}})$. This edit boundary permits introducing} CTE-derived \blue{scopes but still checks that the original tables and filters are subsumed by the inner scope}. In the first LLM attempt, the model aggressively removes \textit{all} original tables $\{\texttt{sets}, \texttt{set\_translations}\}$ without introducing replacements---effectively discarding the query's relational foundation. The \blue{temporal monitor} rejects: the table set shrank without any corresponding CTE or subquery definitions. On retry (temperature adjusted to 0.3 for more conservative generation), the LLM restructures the query using two well-defined CTEs (\texttt{yearly\_sets}, \texttt{language\_counts}), each internally referencing the original tables. The \blue{monitor} accepts \blue{the frame check}: the original tables are \textit{subsumed} within CTE definitions, \blue{not discarded}.

\runinhead{Cross-Layer Convergence.}
However, this CTE restructure uses \texttt{YEAR()}, a function absent from SQLite's dialect configuration ($\Delta_{\text{SQLite}}$), introducing two new $\gamma$-layer violations. Although the \blue{frame-preservation} check passed, the temporal monitor detects that the violation count increased ($|V_{\text{cand}}| = 2 > |V| = 1$), violating \blue{monotonic progress}. The transition is therefore \textbf{rejected}.

Re-prompted with both the structural restructuring guidance and the dialect constraint (replacing \texttt{YEAR()} with the SQLite-compatible \texttt{strftime('\%Y', ...)}), the LLM produces a CTE restructure that resolves the $\beta$-scoping violation and avoids introducing any dialect-incompatible functions. The temporal monitor re-validates: $V(\Phi', \Sigma, \Delta) = \emptyset$. \blue{The progress, edit-boundary, and frame checks} are satisfied, and the transition is \textbf{accepted}. The repair converges in a single accepted transition spanning both error classes ($\beta$ and $\gamma$). An EF pipeline, lacking hierarchical error separation and \blue{checked restructuring}, would risk oscillation between structural and dialect fixes.

\subsubsection{Case 3: The Illusion of Execution Feedback}
Consider a SQLite query generated for the question \textit{``What is the monthly average number of schools that opened in Alameda County under the jurisdiction of the Elementary School District in 1980?''}:

\begin{lstlisting}[language=SQL, basicstyle=\ttfamily\small]
SELECT AVG(COUNT(schools.School)) AS MonthlyAverage 
FROM schools
WHERE schools.DOC = 52 AND schools.OpenDate LIKE '1980%'
GROUP BY MONTH(schools.OpenDate);
\end{lstlisting}
The $\alpha$-layer passes (no syntax errors). However, the query suffers from two orthogonal violations. The $\beta$-layer detects a \textbf{nested aggregation violation} ($R_{\beta,3}$): the expression \texttt{AVG(COUNT(...))} nests one aggregate inside another, which must be unpacked into a subquery. The $\gamma$-layer (under $\Delta_{\text{SQLite}}$) detects a \textbf{dialect conformance violation}: the \texttt{MONTH()} function is unsupported in SQLite and must be rewritten to \texttt{strftime('\%m', ...)}.

Under EF, the SQLite engine aborts at the first structural error, emitting only \texttt{misuse of aggregate function COUNT()}---an opaque message that completely masks the dialect issue. In our empirical test with Qwen2.5-Coder, the EF agent fell into a destructive error-handling cycle: to force compilation against the obscure aggregation error, the LLM blindly removed the required outer \texttt{AVG} operation entirely, converging on \texttt{SELECT COUNT(School) AS MonthlyCount ... GROUP BY strftime(\ldots)}. This highlights a severe \textit{semantic degradation}: \emph{execution succeeded, but the requested quantity changed from an average of counts to a count}.

\runinhead{Single-Pass Convergence.}
\thistool\ extracts both violations simultaneously via a single non-executing static pass. By compiling a \blue{bounded candidate} subspace $\Omega_{\text{cand}}$, it explicitly guides the LLM to lower the \texttt{COUNT} into a valid subquery while preserving the outer \texttt{AVG}, and concurrently \blue{apply the template in $\Delta_{\text{SQLite}}$ that replaces} \texttt{MONTH()} \blue{with} \texttt{strftime}. The \blue{Verify-Before-Apply Temporal Monitor rejects candidates that delete the protected} \texttt{AVG} \blue{aggregation instead of repairing the detected violations}. The output is:

\begin{lstlisting}[language=SQL, basicstyle=\ttfamily\small]
SELECT AVG(school_count) AS MonthlyAverage
FROM (
    SELECT COUNT(schools.School) AS school_count
    FROM schools
    WHERE schools.DOC = 52 AND schools.OpenDate LIKE '1980%'
    GROUP BY strftime('%m', schools.OpenDate)
) AS subquery
\end{lstlisting}
The \blue{accepted repair preserves the statically protected aggregation and filter structure while} resolving both the $\beta$-layer \blue{nested-aggregate} violation and the $\gamma$-layer dialect violation simultaneously. An EF pipeline, discovering errors sequentially through execution, risks destructive semantic regression to achieve superficial compilation success.

\FloatBarrier
\section{Experimental Evaluation}
\label{sec:experiments}

The experimental evaluation is designed to \blue{evaluate the enforcement of the runtime invariants by} our neuro-symbolic monitor and their practical impact on SQL generation.
We investigate the empirical behavior of \thistool\ by answering four Research Questions (RQs):
\begin{itemize}[leftmargin=*, nosep]
    \item \textbf{RQ1 (End-to-End Efficacy \& Universality):} Does the framework universally improve repair execution accuracy across diverse datasets and generation pipelines?
    \item \textbf{RQ2 (\blue{Baseline Comparisons and Runtime Invariants}):} \blue{How do baselines compare with ours, and how does the monitor behave under monotonic progress, regression safety, and bounded retries?}
    \item \textbf{RQ3 (Multi-Dialect Portability):} Can \blue{a fixed} symbolic dialect configuration~$\Delta$ maintain convergence stability across diverse SQL engines (SQLite, MySQL, PostgreSQL) without re-training?
    \item \textbf{RQ4 (Component Necessity \& \blue{Proposer} Universality):} What are the individual contributions of the framework's components, and does the \blue{monitor reliably control SQL repair proposals across diverse foundation} model architectures?
\end{itemize}

\begin{table*}[ht!]
\centering
\resizebox{\textwidth}{!}{%
\footnotesize
\begin{tabular}{cccccccc}
\hline
\multirow{2}{*}{\textbf{Method}} &
\multirow{2}{*}{\textbf{\begin{tabular}[c]{@{}c@{}}Inference\\ Model\end{tabular}}} &
\multirow{2}{*}{\textbf{\begin{tabular}[c]{@{}c@{}}Zero-shot \\ Setting\end{tabular}}} &
\multirow{2}{*}{\textbf{Retrieval}} &
\multicolumn{4}{c}{\textbf{Execution Accuracy (\%)}} \\ \cline{5-8}
  &  &  & & \textbf{Simple} & \textbf{Moderate} & \textbf{Challenging} & \textbf{All} \\ \hline


Gen-SQL~\citep{gensql} & Qwen2.5-Coder-7B  & $\times$ &  \checkmark & 53.4 & 34.9 & 26.9 & 45.3 \\
Gen-SQL~\citep{gensql} & Qwen2.5-Coder-32B  & $\times$ & \checkmark & 61.6 & 42.2 & 35.9 & 53.3 \\
TA-SQL~\citep{tasql} & Qwen2.5-Coder-7B  & \checkmark & $\times$ & 54.70 & 33.41 & 29.66 & 45.89 \\
TA-SQL~\citep{tasql} & Qwen2.5-Coder-32B  & \checkmark  & $\times$ & 66.70 &  49.14 & 46.21 & 59.45 \\
\midrule

Gen-SQL + \thistool & Qwen2.5-Coder-7B  & $\times$ & \checkmark & 
\textbf{63.5}\,$\boldsymbol{\uparrow}$\textsubscript{\small +10.1} & 
\textbf{44.0}\,$\boldsymbol{\uparrow}$\textsubscript{\small +9.1} & 
\textbf{35.2}\,$\boldsymbol{\uparrow}$\textsubscript{\small +8.3} & 
\textbf{54.9}\,$\boldsymbol{\uparrow}$\textsubscript{\small +9.6} \\

Gen-SQL + \thistool & Qwen2.5-Coder-32B & $\times$ & \checkmark & 
\textbf{70.8}\,$\boldsymbol{\uparrow}$\textsubscript{\small +9.2} & 
\textbf{57.1}\,$\boldsymbol{\uparrow}$\textsubscript{\small +14.9} & 
\textbf{51.7}\,$\boldsymbol{\uparrow}$\textsubscript{\small +15.8} & 
\textbf{64.9}\,$\boldsymbol{\uparrow}$\textsubscript{\small +11.6} \\

TA-SQL + \thistool & Qwen2.5-Coder-7B & \checkmark & $\times$ & 
\textbf{64.00}\,$\boldsymbol{\uparrow}$\textsubscript{\small +9.3} & 
\textbf{46.98}\,$\boldsymbol{\uparrow}$\textsubscript{\small +13.57} & 
\textbf{38.62}\,$\boldsymbol{\uparrow}$\textsubscript{\small +8.96} & 
\textbf{56.45}\,$\boldsymbol{\uparrow}$\textsubscript{\small +10.56} \\

TA-SQL + \thistool & Qwen2.5-Coder-32B & \checkmark & $\times$ & 
\textbf{70.49}\,$\boldsymbol{\uparrow}$\textsubscript{\small +3.8} & 
\textbf{56.03}\,$\boldsymbol{\uparrow}$\textsubscript{\small +6.9} & 
\textbf{50.34}\,$\boldsymbol{\uparrow}$\textsubscript{\small +4.1} & 
\textbf{64.21}\,$\boldsymbol{\uparrow}$\textsubscript{\small +4.8} \\

\bottomrule
\end{tabular}%

}
\caption{Execution Accuracy on BIRD Development Dataset \blue{(full-pipeline integration).}}
\label{tab:acc-bird-dev}
\end{table*}

\begin{table*}[ht!]
\centering
\footnotesize
\resizebox{\textwidth}{!}{%
\begin{tabular}{cccccccc}
\hline
\multirow{2}{*}{\textbf{Method}} &
\multirow{2}{*}{\textbf{\begin{tabular}[c]{@{}c@{}}Inference\\ Model\end{tabular}}} &
\multirow{2}{*}{\textbf{\begin{tabular}[c]{@{}c@{}}Zero-shot \\ Setting\end{tabular}}} &
\multirow{2}{*}{\textbf{Retrieval}} &
\multicolumn{4}{c}{\textbf{Execution Accuracy (\%)}} \\ \cline{5-8}
  &  &  & & \textbf{Simple} & \textbf{Moderate} & \textbf{Challenging} & \textbf{All} \\ \hline
Gen-SQL~\citep{gensql} & Qwen2.5-Coder-7B  & $\times$ & \checkmark & 82.6 & 57.5 & 50.0 & 72.1 \\
Gen-SQL~\citep{gensql} & Qwen2.5-Coder-32B  & $\times$ & \checkmark & 82.7 & 58.0 & 50.6 & 72.4 \\
TA-SQL~\citep{tasql} & Qwen2.5-Coder-7B  & \checkmark & $\times$ & 78.0 & 59.8 & 34.7 & 65.1 \\
TA-SQL~\citep{tasql} & Qwen2.5-Coder-32B  & \checkmark & $\times$ & 81.3 & 71.8 & 47.1 & 72.1 \\
\midrule

Gen-SQL + \thistool & Qwen2.5-Coder-7B & $\times$ & \checkmark & 
\textbf{83.8}\,$\boldsymbol{\uparrow}$\textsubscript{\small +1.2} & 
\textbf{60.9}\,$\boldsymbol{\uparrow}$\textsubscript{\small +3.4} & 
\textbf{52.4}\,$\boldsymbol{\uparrow}$\textsubscript{\small +2.4} & 
\textbf{73.9}\,$\boldsymbol{\uparrow}$\textsubscript{\small +1.8} \\

Gen-SQL + \thistool & Qwen2.5-Coder-32B & $\times$  & \checkmark & 
\textbf{87.2}\,$\boldsymbol{\uparrow}$\textsubscript{\small +4.5} & 
\textbf{67.2}\,$\boldsymbol{\uparrow}$\textsubscript{\small +9.2} & 
\textbf{57.1}\,$\boldsymbol{\uparrow}$\textsubscript{\small +6.5} & 
\textbf{77.6}\,$\boldsymbol{\uparrow}$\textsubscript{\small +5.2} \\

TA-SQL + \thistool & Qwen2.5-Coder-7B & \checkmark & $\times$ &
\textbf{79.5}\,$\boldsymbol{\uparrow}$\textsubscript{\small +1.5} & 
\textbf{65.5}\,$\boldsymbol{\uparrow}$\textsubscript{\small +5.7} & 
\textbf{47.1}\,$\boldsymbol{\uparrow}$\textsubscript{\small +12.4} & 
\textbf{69.5}\,$\boldsymbol{\uparrow}$\textsubscript{\small +4.4} \\

TA-SQL + \thistool & Qwen2.5-Coder-32B & \checkmark & $\times$ &
\textbf{83.2}\,$\boldsymbol{\uparrow}$\textsubscript{\small +1.9} & 
\textbf{72.4}\,$\boldsymbol{\uparrow}$\textsubscript{\small +0.6} & 
\textbf{54.7}\,$\boldsymbol{\uparrow}$\textsubscript{\small +7.6} & 
\textbf{75.0}\,$\boldsymbol{\uparrow}$\textsubscript{\small +2.9} \\

\bottomrule
\end{tabular}%
}
\caption{Execution Accuracy on Spider Development Dataset \blue{(full-pipeline integration).}}
\label{tab:acc-spider-dev}

\end{table*}
%


\begin{figure}[htbp]
\centering
\scalebox{0.9}{%
\begin{minipage}[t]{0.58\textwidth}
\begin{table}[H]
    \centering
    \small
    \begin{tabular}{@{}llcc@{}}
    \toprule
    \multirow{2}{*}{\textbf{Metric}} & \multirow{2}{*}{\textbf{Strategy}} & \multicolumn{2}{c}{\textbf{Generation Method}} \\
    \cmidrule(lr){3-4}
    & & \textbf{Gen-SQL} & \textbf{TA-SQL} \\
    \midrule
    \multirow{4}{*}{Accuracy (\%)}
    & Base & 45.31 & 45.89 \\
    & Reflection & 47.07 & 46.54 \\
    & EF & 49.88 & 51.30 \\
    & \textbf{Our} & \textbf{\ourgensql} & \textbf{\ourtasql} \\
    \midrule
    \multirow{4}{*}{Left Errors}
    & Base & 340 & 306 \\
    & Reflection & 291 & 280 \\
    & EF & \blue{84} & \blue{72} \\
    & \textbf{Our} & \textbf{\gensqlerr} & \textbf{\tasqlerr} \\
    \bottomrule
    \end{tabular}
    \caption{Performance Comparison of SOTA Methods on BIRD~\citep{BRID} using Qwen2.5-Coder-7B \blue{(repair-only integration)}. \blue{\thistool\ achieves the highest accuracy and fewest residual errors for both pipelines.}}
    \label{tab:debugger}
    
\end{table}
\end{minipage}%
\hspace{0.02\textwidth}
\begin{minipage}[t]{0.5\textwidth}
\begin{table}[H]
    \centering
    \small
    \begin{tabular}{@{}llcc@{}}
    \toprule
    \multirow{2}{*}{\textbf{Metric}} & 
    \multirow{2}{*}{\textbf{Strategy}} & 
    \multicolumn{2}{c}{\textbf{Generation Method}} \\
    \cmidrule(lr){3-4}
    & & \textbf{Gen-SQL} & \textbf{TA-SQL} \\
    \midrule
    \multirow{2}{*}{Tokens (avg)}
    & EF & 1,867 & 2,409 \\
    & \textbf{Our} & \textbf{1,506} & \textbf{\blue{1,679}} \\
    \midrule
    \multirow{2}{*}{Time (s)}
    & EF & 19.53 & 17.68 \\
    & \textbf{Our} & \textbf{\gensqltime} & \textbf{\tasqltime} \\
    \midrule
    \multirow{2}{*}{Rounds (avg)}
    & EF & 2.00 & 2.81 \\
    & \textbf{Our} & \textbf{1.37} & \textbf{1.18} \\
    \bottomrule
    \end{tabular}
    \caption{\blue{Repair Efficiency on BIRD using Qwen2.5-Coder-7B. \thistool\ uses fewer tokens and repair rounds than EF for both pipelines.}}
    \label{tab:efficiency}
\end{table}
\end{minipage}%
}

\end{figure}

\subsection{Experimental Setup}

\runinhead{Environment and Dataset.}
All experiments are conducted on an Ubuntu 22.04 server equipped with 32-core Intel 5.0GHz CPUs using BF16 precision. We evaluate our framework across three established benchmarks: BIRD~\citep{BRID} (1,534 dev questions across 11 databases covering domains such as education, healthcare, and sports) for real-world schema complexity with simple-to-challenging difficulty levels; Spider~\citep{SPIDER} (1,034 dev questions across 20 databases spanning 138 domains) for cross-domain diversity with syntactically varied SQL patterns; and BIRD Mini-Dev~\citep{BRID} (500 questions selected from BIRD-Dev across 11 databases), which we deploy on both PostgreSQL and MySQL in addition to the default SQLite to test multi-dialect portability. Together, these benchmarks span three SQL dialects and over 3,000 evaluation instances.

\runinhead{Implementation.} The temporal monitor and Hierarchical Analyzer are implemented in roughly 10k lines of code. The Monitor enforces the \blue{runtime checks} described in \S~\ref{sec:repair_loop}, with the \blue{per-state} retry bound set to $k_{\max}=5$ (\blue{Def.~\ref{def:bounded-halting}}). We deploy multiple open-weight models and the Qwen2.5-Coder suite (7B and 32B) as our primary LLM and set the default temperature to 1. We emphasize that our objective is not to brute-force the State-of-the-Art (SOTA) execution accuracy through massive scaling or ensembling, but to demonstrate the \textit{relative improvements} in \blue{repair fidelity and bounded halting} that a Property-Guided framework brings uniformly to established pipelines (Gen-SQL~\citep{gensql} and TA-SQL~\citep{tasql}). Finally, we evaluate with 6 diverse LLM \blue{repair proposers} to validate the framework's robustness and universality. Since EF constitutes the post-generation correction mechanism in current Text-to-SQL systems (as discussed in \S\ref{sec:intro} and \S\ref{sec:related_work}), we adopt it as our primary baseline throughout. Our framework is \textit{independent} of the upstream generation strategy\blue{. We evaluate full-pipeline integration, which includes schema-linking checks before SQL generation (Table~\ref{tab:acc-bird-dev}), and repair-only integration, which replaces the EF loop on generated SQL while preserving all other pipeline components (Table~\ref{tab:debugger}).}

\subsection{RQ1: End-to-End Efficacy \& Universality}

\runinhead*{Setup.}
A defining strength of \thistool\ is its architectural universality: \blue{the analyzer can also check intermediate pipeline outputs against the formalized schema~$\Sigma$. In the full-pipeline setting, it checks table and column references in SQL generation workflow, in addition to repairing generated SQL. This allows invalid references to be corrected before they propagate into the final query.} 

The results are in  Tables~\ref{tab:acc-bird-dev} (BIRD) and~\ref{tab:acc-spider-dev} (Spider), which demonstrate that our framework delivers substantial, consistent accuracy improvements across a wide array of configurations. This improvement holds true regardless of the foundation model scale (7B vs. 32B) or the inference methodology (zero-shot, few-shot, with or without retrieval). Specifically, when augmenting Gen-SQL~\citep{gensql} and TA-SQL~\citep{tasql} on the uniquely challenging BIRD dataset, \thistool\ drives accuracy gains of up to +11.6 and +10.56 \blue{percentage points}, respectively. We observe similarly robust gains on the Spider dataset. 
This performance gap between baseline execution feedback and our method yields a critical insight: \textit{integrated hierarchical validation \blue{can outperform terminal} correction}. By preventing error propagation rather than merely reacting to execution failures, \thistool\ translates \blue{monitor-checked repair fidelity} into measurable execution accuracy gains. 


%
\begin{remark}[On the Relationship Between Repair and Accuracy]
\label{rem:accuracy}
Execution accuracy requires two independent conditions: (1)~the SQL must be structurally \textit{executable}, and (2)~the result must \textit{match} the gold answer. Our framework targets condition~(1): we reduce execution failures from \blue{72} to just 16 (a \blue{$77.8\%$} reduction over EF), but these \blue{56} additionally repaired queries are a small fraction of the full benchmark, so the absolute accuracy gain appears moderate. This is by design, \blue{an unexecutable query is always wrong, but an executable one is not necessarily correct}. Our contribution is \textit{repair fidelity}: fixing strictly more bugs while \blue{preserving monitor-protected structure}.
\end{remark}

\subsection{\blue{RQ2: Baseline Comparisons and Runtime Invariants}}
\label{sec:exp_termination}

\begin{table*}[t]
\centering
\footnotesize
\begin{tabular}{@{}llrrrr@{}}
\toprule

\textbf{Strategy} & \textbf{Source} & \textbf{Acc. (\%)} & \textbf{Accepted} & \textbf{Residual} & \textbf{Empty} \\
\midrule
Constrained decoding & Gen-SQL & 50.78  & 204 & 136 & 108 \\
Constrained decoding & TA-SQL & 51.43  & 210 & 96 & 100 \\
\midrule
EF+Rule-Skill & Gen-SQL & 51.89  & 193 & 147 & 108 \\
EF+Rule-Skill & TA-SQL & 51.30  & 184 & 122 & 94 \\
\midrule
\thistool & Gen-SQL & \textbf{\ourgensql} & \textbf{296} & \textbf{\gensqlerr} & 115 \\
\thistool & TA-SQL & \textbf{\ourtasql} & \textbf{290} & \textbf{\tasqlerr} & 82 \\

\bottomrule
\end{tabular}
\caption{\blue{Comparison with constrained decoding and EF+Rule-Skill on BIRD. \thistool\ achieves higher accuracy and fewer residual errors than both baselines for both pipelines.}}
\label{tab:additional-baselines}

\end{table*}

\thistool\ rests on two
\blue{monitored runtime invariants: bounded soundness with respect to the declared rule family and bounded halting under the finite retry budget. These invariants are narrower than semantic correctness.}

\blue{\runinhead{Baseline Comparisons.} Table~\ref{tab:additional-baselines} compares \thistool\ with EF+Rule-Skill and a constrained-decoding baseline using SQL syntax and schema checks, which follow~\citet{scholak-etal-2021-picard,poesia2022synchromesh} works. The constrained-decoding baseline validates complete SQL candidates and accepts them only after successful SQL execution. All methods operate on outputs from unchanged Gen-SQL and TA-SQL pipelines. \thistool\ achieves higher accuracy and fewer residual errors for both pipelines.}

\begin{table}[tbp]
\centering

\scalebox{0.95}{
\footnotesize
\begin{tabular}{@{}l*{4}{c}@{}}
\toprule
\textbf{Error Type} & \textbf{Reflection} & \textbf{EF} & \textbf{Our-7B} & \textbf{Our-32B} \\
\midrule
Syntactic          & 61 & 0 & \textbf{0}   & \textbf{0}  \\
Context-Free       & 19 & 8 & \textbf{4}   & \textbf{2}  \\
Context-Sensitive  & 200 & 64 & \textbf{\blue{12}} & \textbf{\blue{5}}  \\
\bottomrule
\end{tabular}
}
\caption{Residual Error Analysis (TA-SQL on BIRD)\blue{: both model sizes leave fewer context-free and context-sensitive errors than EF}.}
\label{tab:error_analysis}

\end{table}
\runinhead{Canonical Error Resolution.} Table~\ref{tab:error_analysis} tracks the residual errors across all layers\blue{, totaling 16 and 7 for the 7B and 32B models, respectively}. While basic Execution Feedback (EF) struggles with Context-Sensitive ($\gamma$) errors---leaving 64 unresolved violations---our property extraction reduces these by \blue{81.25\% with the 7B model}. As summarized in Table~\ref{tab:debugger}, this determinism minimizes the total number of unresolved errors \blue{and blocks accepted fix-and-regress transitions.} A concrete illustration is given in Case~3 (\S\ref{sec:case_studies}), where EF silently corrupts the user's aggregation \blue{structure} to achieve superficial compilation success---precisely the kind of \blue{protected structural} degradation that our monitor rejects.


\begin{table}[tbp]
\centering
\vspace{2mm}
\footnotesize
\begin{tabular}{@{}l*{5}{c}@{}}
\toprule
\textbf{Metric} & \textbf{TA-SQL} & \textbf{EF} & \textbf{Align-LLM} & \blue{\textbf{Align+Reject}} & \textbf{Align-only} \\
\midrule
Accuracy (\%)           & 45.89 & 46.09 & 45.96 & \blue{46.41} & 46.09 \\
Empty Results           & 85    & 73    & 57    & \blue{69} & 82    \\
New Violations   & --    & 6      & 21    & \blue{\textbf{0}} & \textbf{0} \\
Time (s)                & --    & 1500   & 1046  & \blue{1090} & \textbf{45}    \\
\bottomrule
\end{tabular}
\caption{\blue{Soundness when handling empty results.} Both EF and Alignment-LLM suffer regressions on BIRD dataset. Align is \detalign\ and Align+Reject is Alignment-LLM+rejection.}
\label{tab:alignment}

\end{table}
\blue{\runinhead{Accepted Regression Blocking.} The empty-result study} (Table~\ref{tab:alignment}) \blue{isolates where safety comes from: the unconstrained Alignment-LLM introduces new static violations, but gating the same LLM with monitor-style rejection yields no new violations in this experiment, supporting the use of an acceptance gate rather than relying on the generator alone. The} deterministic $\omega_{\text{align}}$ operator \blue{preserves this gate-enforced safety at near-identical accuracy while invoking no LLM; given the latency--accuracy balance we adopt it in the framework rather than the LLM variant, which buys marginal accuracy at an order-of-magnitude latency cost and reintroduces the nondeterminism our analysis eliminates. In the core} loop, across 308 repaired queries, the \blue{Verify-Before-Apply Temporal Monitor} intercepted up to 77 invariant violations and enforced 9 \blue{finite-budget stops.}

\begin{figure*}[t]
    \centering
    \includegraphics[width=0.95\textwidth]{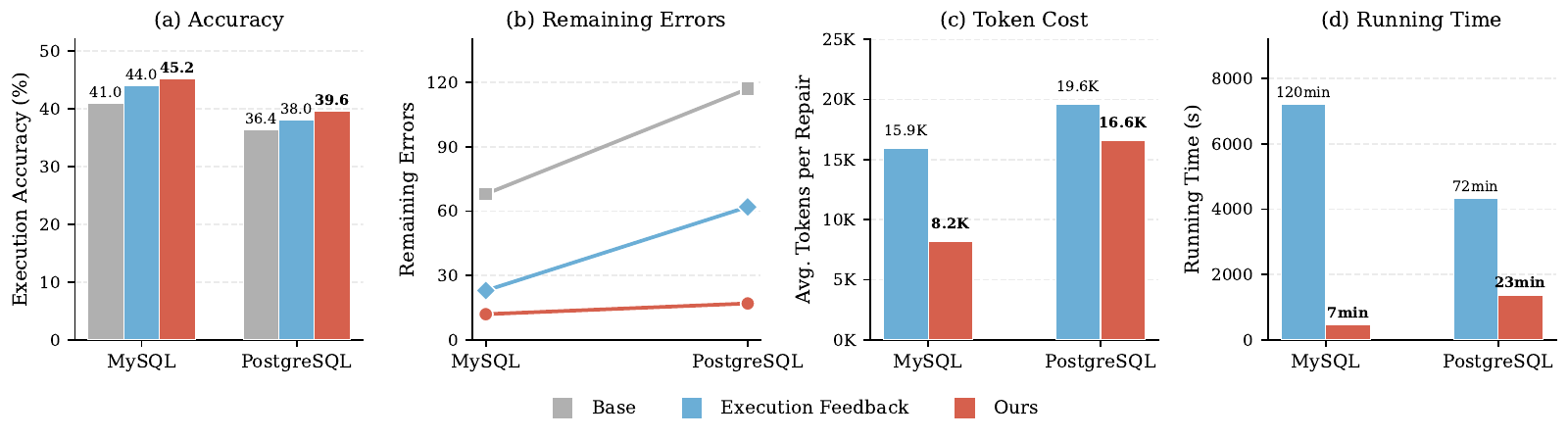}
    \caption{Multi-dialect Convergence. (a) Performance across dialects, (b) Remaining errors after repair, (c) Token consumption comparison, (d) Execution time comparison.}
    \label{fig:multi-dialect}
    
\end{figure*}


{\runinhead{\blue{Enforcing Bounded Halting.}} 
By strictly pruning cyclic fixes via the monitor, we observe dramatic efficiency gains. As detailed in Table~\ref{tab:efficiency}, \blue{\thistool\ reduces average repair rounds by 58.0\% for TA-SQL}. Convergence is highly concentrated: 98.3\% of TA-SQL and 93.79\% of \blue{Gen-SQL} repaired queries converge within the first two rounds; the remaining queries either hit the retry cap $k_{\max}$ or are conservatively abandoned by the Monitor. Across 237 LLM-invoked repair sessions, \thistool\ maintained an average consumption of just 1,679 tokens per query (totaling $\sim$\blue{398k} tokens), reducing \blue{average token consumption by approximately} $30\%$ over EF. This highlights an architectural advantage: \thistool\ \blue{localizes token consumption around rule-indexed operator schemas rather than asking} the LLM to \blue{rediscover root causes from execution traces}. Case~3 (\S\ref{sec:case_studies}) demonstrates this concretely: while the EF agent requires multiple rounds and ultimately degrades \blue{protected aggregation structure} to achieve compilation, \thistool\ resolves both the $\beta$ and $\gamma$ violations in a single unexecuted \blue{accepted transition}.}

\subsection{RQ3: Multi-Dialect Portability}
\label{sec:exp_dialect}

{\runinhead{Dialect-Aware Convergence Strategy.} 
Despite the amplified complexity of stricter dialects, the dialect-aware configuration $\Delta$ effectively adapts the static analyzer's $\gamma$-layer validation rules to handle dialect-specific constraints \textit{before} neural invocation. As depicted in Figure~\ref{fig:multi-dialect}(c--d), the EF baseline struggles profoundly under PostgreSQL's strictness, consuming approximately 2.3M tokens across 239 repair iterations as it attempts to blindly guess type casts and division guards. In contrast, \thistool\ achieves a higher successful repair rate (fixing 85.5\% of execution errors) while consuming approximately 1.9M tokens. On MySQL, the advantage is equally clear: \thistool\ resolves 82.4\% of execution failures at roughly half the token cost of EF (556k to 1.08M tokens).
\blue{Crucially, while EF relies on autoregressive generation loops that can spiral when faced with cascading type errors, \thistool\ deterministically guards against these violations via static analysis.} By preemptively isolating schema mismatches and syntactic incompatibilities at the $\gamma$-layer, our framework truncates the deep cascading error chains that defeat pure execution feedback, maintaining structural convergence across both permissive and strict database configurations.}

{\runinhead{Dialect-Specific Challenges.}
Our cross-dialect experiment reveals a deeper challenge rooted in type-system strictness. When executing identical queries generated by the same model, MySQL produces 68 execution exceptions, whereas PostgreSQL yields 117 errors, representing a 72\% increase on the same natural-language workload (Figure~\ref{fig:multi-dialect}a-b). We analyze three mechanistic drivers for this disparity. 
First, PostgreSQL enforces strict typing and rejects the implicit coercions that MySQL routinely tolerates. For example, applying the \texttt{ROUND} function to \blue{a double precision first argument} triggers 70 explicit-cast and type-incompatibility diagnostics in PostgreSQL that are entirely absent in MySQL. 
Second, the engines differ fundamentally in error handling for undefined operations. While MySQL returns \texttt{NULL} upon division by zero, PostgreSQL terminates execution, contributing to 91 division-related diagnostics compared to only 16 in MySQL.
Finally, identifier resolution rules exacerbate schema mismatches. PostgreSQL's automatic case-folding of unquoted identifiers causes LLM-generated CamelCase references to fail static resolution.
Cumulatively, these constraints generate a cascading effect where a single query on PostgreSQL triggers complex multi-error failures, demanding significantly more intricate repair trajectories. Case~1 (\S\ref{sec:case_studies}) provides a concrete illustration: a single PostgreSQL query simultaneously triggers both a type-cast violation (\texttt{ROUND} applied to \texttt{FLOAT} instead of \texttt{NUMERIC}) and an identifier binding failure (\texttt{customerid} vs.\ the schema-required \texttt{CustomerID}), neither of which would surface on MySQL.}


\subsection{RQ4: Component Necessity and \blue{Proposer} Universality}

%

\begin{table*}[!t]
\centering
\footnotesize
\begin{tabular}{@{} lccc @{}}
\toprule
\textbf{Configuration} & \textbf{Accuracy (\%)} & \textbf{Errors Left} & \textbf{Tokens (avg)} \\
\midrule
Baseline (TA-SQL-7B Generated) & 45.89 & 306 & -- \\
\midrule
\thistool\ (Full System) & \textbf{\ourtasql} & \textbf{\tasqlerr} & \textbf{1,679} \\
- \blue{w/o \detalign} & \blue{52.67} & \blue{16} & \blue{1,679} \\
- w/o Verify-Before-Apply Temporal Monitor & 52.00 & 25 & 1,902 \\
- w/o Hierarchical Validation & 52.26 & 39 & 2,426 \\
- w/o Plan Synthesis & 50.24 & 52 & 2,916 \\

\bottomrule
\end{tabular}
\caption{Ablation Study: Component Contributions to \thistool\ Performance.}
\label{tab:ablation}

\end{table*}

\runinhead{Component Necessity.} Table~\ref{tab:ablation} isolates the contributions of the core components. Removing the temporal monitor increases residual errors from \tasqlerr\ to 25 and slightly raises token usage (1,679$\to$1,902), confirming its role in preventing regressive patches that would otherwise undo prior fixes. Removing Hierarchical Validation is more damaging: errors climb to 39 and tokens rise to 2,426, as the LLM must compensate for the absent structural decomposition with broader speculation. Removing Plan Synthesis has the most severe impact: errors surge to 52 and tokens to 2,916, confirming that formal constraints actively truncate the LLM's hypothesis space, yielding measurable efficiency gains. \blue{Removing \detalign\ costs 0.2 accuracy points (\ourtasql$\to$52.67) with residual execution errors unchanged (16), matching its isolated behavior in Table~\ref{tab:alignment}.}

%

\Needspace{14\baselineskip} 
\begin{table}[tbp]
\vspace{2mm}
\centering
\footnotesize

\begin{tabular}{@{}llcc@{}}
\toprule
\textbf{Rule} & \textbf{Property} & \textbf{SQLite} & \textbf{LOO $\Delta$} \\
\midrule
$R_{\alpha,\pi}$ & parse/syntax        & 21  & $\downarrow\,1.2$ \\
$R_{\beta,1}$    & GROUP BY scope      & 1   & $\downarrow\,0.2$ \\
$R_{\beta,2}$    & ORDER BY scope      & 1   & $\downarrow\,0.2$ \\
$R_{\beta,3}$    & nested aggregate    & 4   & $\downarrow\,0.6$ \\
$R_{\gamma,1}$   & identifier resolution & 239 & $\downarrow\,5.1$ \\
$R_{\gamma,2}$   & FK connectivity     & 71  & $\downarrow\,2.6$ \\
$R_{\gamma,3}$   & dialect type        & 18  & $\downarrow\,0.5$ \\
\bottomrule
\end{tabular}
\caption{\blue{Rule triggers and leave-one-rule-out performance drops (LOO $\Delta$) on BIRD (SQLite). Identifier resolution has the largest drop.}}
\label{tab:rule-counts}

\end{table}
\blue{\runinhead{Per-Rule Generality.} {The rule family is open and extensible (Remark~\ref{rem:operator_bounds}), so rather than enumerate it we report how often each representative rule of Table~\ref{tab:validation-rules} fires. As shown in Table~\ref{tab:rule-counts}, every rule fires on real violations and the family collectively covers all logged errors---detection coverage, not accuracy magnitude, is the contribution we claim. A leave-one-out check confirms each rule contributes, but because one query typically triggers several interacting rules, the accuracy deltas are entangled and should not be summed.}}

{\runinhead{\blue{Proposer} Universality and Non-Compliance Robustness.} \blue{The monitor's acceptance checks are independent of the repair proposer's architecture.} Table~\ref{tab:compliance} \blue{evaluates repair performance and non-compliance interception} across \blue{six model configurations} spanning State Space Models (Codestral), Mixture of Experts (\blue{gpt-oss-120B}), and dense Transformers of varying scale (Qwen-7B/32B, Llama-70B, \blue{Mistral-24B}). Every model shows consistent accuracy gains over EF, ranging from $+1.43$ to $+6.13$ \blue{percentage points}. This variance reflects differences in how cleanly each upstream generator produces SQL---models that already generate fewer structural errors leave less room for our repair layer to improve absolute accuracy (see Remark~\ref{rem:accuracy}).  Critically, the \textit{Non-compliance} row shows that LLMs routinely violate repair specifications (up to 15 out-of-bounds patches for larger models), yet \blue{\thistool} intercepts 100\% of these \blue{observed non-compliant proposals. These results support the monitor's ability to enforce the declared repair contracts across different proposers.}}

\begin{table}[t]
\centering
\footnotesize
\scalebox{0.95}{
\begin{tabular}{lcccccc}
\toprule 
\textbf{Metric} & \textbf{Qwen2.5-7B} & \textbf{Qwen2.5-32B} & \textbf{Codestral-7B} & \textbf{gpt-oss-120B} & \textbf{{Llama-3.1-70B}} & \textbf{{Mistral-24B}}\\
\midrule
{EF Accuracy (\%)}   & {51.30}    &   {52.00}  & {51.37}      & {48.04} & {48.37}   & {48.24}  \\
{Our Accuracy (\%)}   & {\ourtasql}    &   53.59   &  {52.80}     & {53.19} & {53.89} & {54.37}   \\
{Non-compliance}   & {2}    & {1}     & {2}      & {15}  & {15}  & {1} \\
{Capture Rate}   & {100\%} & {100\%} & {100\%} & {100\%} & {100\%} & {100\%} \\
\bottomrule
\end{tabular}
}
\caption{Robustness to Instruction Non-Compliance\blue{: across all six proposers, \thistool\ improves accuracy over EF and the monitor intercepts all observed non-compliant proposals}.}

\label{tab:compliance}
\end{table}

\section{Discussion}
\label{sec:discussion}

\runinhead{Scope of Formal Guarantees.}
Our soundness and bounded \blue{halting} guarantees are formally bounded by the properties defined in our $\alpha, \beta, \gamma$ hierarchy. A query that passes all three layers may still be \textit{semantically incorrect} (e.g., selecting the wrong aggregation function). Our guarantees \blue{prevent resolved violations from recurring and ensure bounded halting}; they do not replace oracle-guided semantic verification. The \detalign\ operator (\S\ref{sec:delta_layer}) is explicitly heuristic: it queries data-level evidence rather than static properties, and may conservatively alter filter predicates to resolve empty results. We isolate it as an optional post-validation step.

\runinhead{Generalization and Dialect Extension.}
Our experiments cover SQLite, MySQL, and PostgreSQL. Extending to additional dialects requires only adding a constant-size declarative YAML configuration that specifies dialect-specific operator signatures and AST node equivalences (Remark~\ref{rem:dialect_cost}), with no retraining or modification of the core verification engine.

\runinhead{Evaluation Methodology.}
We adopt execution accuracy following standard BIRD and Spider protocols, complemented by residual error counts across all three semantic layers as a direct structural quality measure (see Remark~\ref{rem:accuracy}).

\runinhead{Residual Limitations.}
Repairing structurally complex SQL remains an open challenge: SQL's own design deficiencies~\citep{shute2024sql} give rise to entangled constructs that resist modular decomposition. As formalized in Remark~\ref{rem:operator_bounds}, our repair completeness is bounded by the $\alpha, \beta, \gamma$ rule coverage; constructs outside this boundary are conservatively left untouched rather than speculatively modified. \blue{Among the 16 residual BIRD errors, approximately 6--8 require constructs outside $\mathcal{L}_{\mathrm{sql}}$ (a fundamental scope boundary), 3--4 are within $\mathcal{L}_{\mathrm{sql}}$ but uncovered by the current declared rules (an extensible rule gap), and 4--6 are LLM instruction non-compliance cases that are intercepted by the monitor rather than accepted as regressions.} Beyond rule coverage, the capability of the underlying LLM \blue{proposer} is a limiting factor: scaling from Qwen2.5-Coder-7B to 32B reduces residual errors from \blue{16 to 7} (see Table~\ref{tab:error_analysis}), indicating that stronger models better follow the constrained repair specifications.

\section{Related Work}
\label{sec:related_work}
\runinhead{Program Repair and Static Analysis.}
Automated program repair~\citep{apr_survey, apr_survey2} spans search-based~\citep{GenProg}\blue{, }constraint-based~\citep{SAVER, SemFix, mechtaev2016angelix, Footpatch}, and template-driven\blue{~\citep{liu2023template} approaches, and a growing line of work couples LLMs with static analyzers, compilers, and linters as deterministic feedback channels for generated code, such as integrating analyzer knowledge into code review~\citep{jaoua2025combininglargelanguagemodels}. Narrowing to SQL}, early rule-based \blue{repair~\citep{presler2021sqlrepair} corrects only }limited syntactic errors\blue{, while SQL static analysis ranges from verifying }generated query strings~\citep{gould2004static} to proving query equivalence\blue{~\citep{wang2024verieql} and industrial linting~\citep{sqlfluff}; property-guided }repair~\citep{ProveNFix} \blue{further checks repair transitions against explicit properties. These approaches, however, treat analyzers as external feedback providers rather than defining a canonical intermediate representation, rule-indexed violation identities, and a bounded monitor over accepted LLM }repair transitions. \thistool\ \blue{is, to our knowledge, the first Text-to-SQL repair framework to combine canonicalization, rule-indexed violation identities, candidate-space repair, and monitor-checked accepted transitions}.

\runinhead{Neuro-Symbolic Generation and Constrained Decoding.}
Constrained decoding methods~\citep{poesia2022synchromesh, mundler2025type, scholak-etal-2021-picard} enforce correctness by restricting the LLM's output at the token level during generation. PICARD~\citep{scholak-etal-2021-picard}, for instance, filters infeasible tokens against a partial SQL grammar and can validate table and column names, effectively handling syntactic errors (our $\alpha$-layer) and partial identifier binding. However, these methods cannot address context-sensitive consistency errors (e.g., cross-table join integrity, dialect-specific type constraints) whose root causes are structurally distant from their token-level manifestation. Moreover, constrained decoding requires white-box access to the model's logit stream, tying it to a specific architecture. 
In contrast, our \blue{Verify-Before-Apply Temporal Monitor} is fully model-agnostic: the LLM proposes candidate patches as a black box, which our engine deterministically verifies against the extracted properties. This decoupling enables universal deployment across diverse LLM architectures.

\runinhead{Text-to-SQL Generation and Post-Processing.}
LLM-based Text-to-SQL systems~\citep{alphasql,guo-etal-2019-towards, scholak-etal-2021-picard, li2023graphixt5mixingpretrainedtransformers, sun2023sqlpalm, SQLFixAgent,gensql, tasql,chasesql,sqlo1,dinsql,macsql,magic,dailsql,opensql,xiyansql} have achieved increasingly strong performance on standard benchmarks through advances in schema linking, reasoning, and scaling. Nevertheless, any generation-based approach inevitably produces structural and schema-level errors due to the non-deterministic nature of LLM decoding. The dominant post-processing paradigm relies on Database Execution Feedback~(EF)~\citep{sun2023sqlpalm, deng2024sqlcraft, llmfixcode, SQLFixAgent,dinsql,sqlfix_template1,sqlo1,gensql}, which provides only coarse-grained dynamic symptoms rather than precise root-cause analysis, forcing LLMs to speculate on fixes and often leading to cyclic failures~\citep{chen2023selfdebug, huang2024largelanguagemodelsselfcorrect}. Multi-agent frameworks~\citep{SQLFixAgent,magsql,magic, chaturvedi2025sqlofthoughtmultiagentictexttosqlguided} and self-correction strategies~\citep{madaan2023selfrefine, magic} attempt to mitigate this through iterative refinement, but lack formal convergence guarantees: without explicit property specifications, each correction step may introduce new regressions. Our framework replaces speculative EF with strict property analysis, providing \blue{bounded runtime-invariant checks and bounded halting} over the repair loop.

\section{Conclusion}

We presented \thistool, a neuro-symbolic framework that recasts Text-to-SQL repair as property-guided program repair. By interposing a formal layer---dialect-aware canonicalization, hierarchical static analysis \blue{under the stratified dependency order $\alpha \prec_{\mathrm{sd}} \beta \prec_{\mathrm{sd}} \gamma$}, and a \blue{verify-before-apply} temporal monitor with \blue{bounded runtime invariants---between} the neural generator and the accepted output, \thistool\ breaks the Arbitration Paradox without sacrificing the flexibility of LLM-based synthesis.
Empirically, \thistool\ resolves \blue{77.8\%} of the execution errors that EF leaves unrepaired, reduces repair rounds by over 58\%, and \blue{\emph{accepts no statically detectable structural regressions in our evaluation}} across three SQL dialects and six LLMs. As a generator-agnostic post-processing layer, it integrates into any upstream pipeline as a drop-in replacement for the EF paradigm. \blue{These runtime invariants} are bounded by the \blue{declared $\alpha,\beta,\gamma$ rules and operator edit-boundary checks.} Enriching the hierarchy with data-aware properties such as cardinality bounds or functional-dependency constraints is a natural next step.

\FloatBarrier
\bibliographystyle{plainnat}
\bibliography{refs}
\clearpage
\appendix
\section{Repair Prompts and Case Study}\label{app:repair-prompts}

This case expands the motivating example in \S~2, showing the system and user prompts, SQL candidates, and decisions of the \thistool.

\subsection{Case setup}
Query:
\begin{quote}
\emph{What is the district ID for the account associated with order 33333?}
\end{quote}
The generator produces:
\begin{lstlisting}[style=sql]
SELECT account.district_id FROM order WHERE order_id = 33333;
\end{lstlisting}
The table name \texttt{order} needs escaping, and \texttt{account.district\_id} references a table absent from the query's sources. Deterministic keyword escaping first changes \texttt{order} to \texttt{`order`}. The two LLM rounds below start from this normalized query, with one remaining binding violation.

\subsection*{System prompts}
\begin{lstlisting}[style=prompt]
You are an experienced SQLite database expert.
Repair SQL using the supplied database schema, error diagnostics,
and repair instructions. Follow the provided constraints and
return a complete SQL query in SQLite syntax.
\end{lstlisting}

\subsection*{Repair boundary in this case}
The JOIN repair may add a source table and its join predicate, corresponding to $\rho.T$ and $\rho.J$ in the CQS. It must preserve the projected attribute \texttt{account.district\_id} and the filter \texttt{order\_id = 33333}. Qualifying that filter as \texttt{`order`.order\_id = 33333} preserves its canonical reference.

\newpage
\subsection{Round 1}
The binding diagnostic identifies \texttt{account} as the missing source table. The prompt supplies a concrete JOIN path and states which existing query parts must be preserved.

\textbf{Representative user message}
\begin{lstlisting}[style=prompt]
# Database dialect
SQLite

# Schema excerpt
account(account_id, district_id)
order(order_id, account_id)
Foreign key: order.account_id -> account.account_id

# Question
What is the district ID for the account associated with order 33333?

# Current SQL
SELECT account.district_id
FROM `order`
WHERE order_id = 33333;

# Focused diagnosis
Error type: COLUMN_NOT_FOUND_NEEDS_JOIN
Target: account.district_id
The column exists in account, but account is not a query source.
Required action: ADD_JOIN

# Suggested JOIN path
JOIN account ON `order`.account_id = account.account_id

# Repair instructions
1. Add account to the query using the suggested JOIN path.
2. Use only tables and columns present in the supplied schema.
3. Keep account.district_id as the projected attribute.
4. Preserve the filter order_id = 33333 and its constant value.
5. Do not add unrelated predicates or remove existing query clauses.
6. Keep the reserved table name `order` escaped.

Apply the repair and output the complete corrected SQL in a SQL code block.
\end{lstlisting}

\textbf{Candidate from the motivating example}
\begin{lstlisting}[style=sql]
SELECT account.district_id
FROM `order`
JOIN account
    ON `order`.account_id = account.account_id;
\end{lstlisting}

\textbf{Monitor decision: rejected.}

\newpage
\subsection{Round 2}
The schema, current SQL, diagnosis, and JOIN path remain available. The new information is the rejection reason and an explicit instruction to retain the filter that the first candidate removed.

\textbf{Representative user message}
\begin{lstlisting}[style=prompt]
# Database dialect
SQLite

# Schema excerpt
account(account_id, district_id)
order(order_id, account_id)
Foreign key: order.account_id -> account.account_id

# Current SQL -- the rejected candidate was not applied
SELECT account.district_id
FROM `order`
WHERE order_id = 33333;

# Focused diagnosis
Error type: COLUMN_NOT_FOUND_NEEDS_JOIN
Target: account.district_id
Add the missing source table account using this JOIN path:
JOIN account ON `order`.account_id = account.account_id

# Previous verification result
Your last repair was REJECTED because it removed the WHERE clause.
Failed check: preservation of the existing filter.
The permitted JOIN repair does not authorize removing this condition:
    order_id = 33333

# Instructions for this retry
1. Start from the current SQL above and add the suggested JOIN.
2. Keep account.district_id in SELECT.
3. Keep WHERE order_id = 33333. Do not change 33333.
   You may qualify the column as `order`.order_id.
4. Keep the original source table and use only schema-valid identifiers.
5. Do not remove clauses or introduce unrelated filters.
6. Keep the reserved table name `order` escaped.

Do not repeat the rejected edit. Output the complete corrected SQL
in a SQL code block, including the preserved WHERE clause.
\end{lstlisting}

\textbf{Candidate from the motivating example}
\begin{lstlisting}[style=sql]
SELECT account.district_id
FROM `order`
JOIN account
    ON `order`.account_id = account.account_id
WHERE `order`.order_id = 33333;
\end{lstlisting}

\textbf{Monitor decision: accepted.}

\clearpage
\subsection{System prompt templates}
The templates below summarize the information supplied to the repair proposer. Angle-bracket placeholders, such as \texttt{\textless CURRENT\_SQL\textgreater}, are filled with the current query, schema, diagnostics, or candidate operators. Optional fields are omitted when unavailable. The diagnostic blocks specialize the common request to the active repair operators. The Verify-Before-Apply Temporal Monitor checks the returned candidate before applying it.

\subsection*{System prompts}
\begin{lstlisting}[style=prompt]
You are an experienced <SQL_DIALECT> database expert. 
Repair SQL using the supplied schema, diagnostics, and candidates.
Follow the permitted edits and preservation requirements. 
Use only schema-valid identifiers and <SQL_DIALECT> syntax. 
Return one complete SQL query in a SQL code block.
\end{lstlisting}

\subsection*{P1. Common repair request}
This request collects the inputs to \textsc{BuildSpec} in Algorithm~2. P2--P7 specialize its diagnostic and repair-instruction fields, P8 adds rejection feedback, and P9 illustrates a request with multiple diagnostics.
\begin{lstlisting}[style=prompt]
# Database dialect
<SQL_DIALECT>

# Database schema
<TABLES_COLUMNS_TYPES_PRIMARY_KEYS_AND_FOREIGN_KEYS>

# Question, when available
<NATURAL_LANGUAGE_QUESTION>

# Current SQL
<CURRENT_SQL>

# Diagnostics and admissible repair candidates
<FOCUSED_DIAGNOSTICS_AND_OPERATOR_SPECIFIC_INSTRUCTIONS>

# Operator edit boundary
<PERMITTED_EDITS_FOR_THE_ACTIVE_OPERATORS>

# Preservation requirements
<QUERY_COMPONENTS_AND_EXPRESSIONS_TO_PRESERVE>

# Resolved-violation history
<PREVIOUSLY_RESOLVED_VIOLATIONS_OR_NONE>

# Rejection feedback, when present
<FAILED_MONITOR_CHECKS_AND_REPAIR_GUIDANCE>

# Task
Repair the reported violations using the supplied candidates.
Respect the edit boundary and preservation requirements.
Do not reintroduce a previously resolved violation.
Output the complete corrected SQL in a SQL code block.
\end{lstlisting}

\clearpage
\subsection*{P2. Identifier binding}
This block instantiates the replacement operator when an existing query source provides an admissible replacement for an unresolved column reference. For missing source tables, use P3.
\begin{lstlisting}[style=prompt]
# Binding diagnostic
Reference: <INVALID_REFERENCE>
Scope: <TARGET_QUERY_SCOPE>
Reason: <BINDING_DIAGNOSTIC>

# Admissible replacements
<REFERENCES_FROM_CURRENT_SOURCES_OR_SCOPE_EXPORTS>

# Required edit
Replace <INVALID_REFERENCE> with <SELECTED_REPLACEMENT>.
Apply the replacement within <TARGET_QUERY_SCOPE>.
Follow the dialect's identifier naming and quoting requirements.

# Preserve
Keep the existing query sources, JOIN predicates, filters,
aggregations, ordering, and non-target expressions unchanged.
Do not add a table or restructure the query for this replacement.
Do not invent a reference outside the supplied candidates.
\end{lstlisting}

\subsection*{P3. Missing source table or FK connection}
The prompt supplies candidate paths derived from the schema's FK graph. The authorized changes are the source-table and JOIN fields of the affected query scope.
\begin{lstlisting}[style=prompt]
# JOIN diagnostic
Scope: <TARGET_QUERY_SCOPE>
Missing source or disconnected tables: <REPORTED_TABLES>
Affected reference, if any: <UNRESOLVED_REFERENCE>
Reason: <JOIN_OR_BINDING_DIAGNOSTIC>

# Candidate FK paths
1. <FK_JOIN_PATH_1>
2. <FK_JOIN_PATH_2_IF_AVAILABLE>

# Required edit
Use <SELECTED_FK_JOIN_PATH> to connect the required tables.
Add only the source or bridge tables and JOIN conditions
authorized by that path, within <TARGET_QUERY_SCOPE>.
Use the supplied schema-valid key columns.

# Preserve
Keep the projection, filters and their literal values,
GROUP BY, HAVING, ORDER BY, and LIMIT unchanged.
Retain the original query sources.
Do not delete the affected reference to hide the binding error.
\end{lstlisting}

\clearpage
\subsection*{P4. Dialect-specific type repair}
For the cast operator, the diagnostic identifies the offending operand and the dialect-valid target types. The prompt asks for a local cast rather than replacement of the surrounding computation.
\begin{lstlisting}[style=prompt]
# Type diagnostic
Dialect: <SQL_DIALECT>
Operator and argument position: <OPERATOR_AND_ARGUMENT_POSITION>
Operand: <OFFENDING_OPERAND_EXPRESSION>
Observed type: <OBSERVED_TYPE>
Required signature: <DIALECT_VALID_OPERATOR_SIGNATURE>

# Admissible casts
<ALLOWED_TARGET_TYPES_AND_CAST_CANDIDATES>

# Required edit
Apply <SELECTED_CAST> to <OFFENDING_OPERAND_EXPRESSION>.
Keep the operand expression inside the cast unchanged.
Use the cast syntax prescribed for <SQL_DIALECT>.

# Preserve
Keep the enclosing computation and non-target arguments.
Preserve the query's tables, joins, filters, grouping, and ordering.
Do not remove the failing expression or change its aggregate
function merely to avoid the reported type error.
\end{lstlisting}

\subsection*{P5. GROUP BY scope repair}
This block follows the declared \texttt{add\_group} operator: it adds missing grouping attributes while preserving the other CQS components.
\begin{lstlisting}[style=prompt]
# Grouping diagnostic
Scope: <TARGET_QUERY_SCOPE>
Aggregation context: <AGGREGATION_CONTEXT>
Ungrouped non-aggregate references: <REPORTED_REFERENCES>
Current GROUP BY: <CURRENT_GROUPING_ATTRIBUTES>

# Admissible additions
<GROUPING_ATTRIBUTES_SELECTED_BY_THE_REPAIR_PLAN>

# Required edit
Add <SELECTED_GROUPING_ATTRIBUTES> to GROUP BY.
Retain the existing grouping attributes.
Restrict this edit to the reported query scope.

# Preserve
Keep the SELECT expressions and aggregate functions unchanged.
Preserve FROM, JOIN, WHERE, HAVING, ORDER BY, and LIMIT.
Do not remove a projected expression or wrap it in an unrelated
aggregate to bypass the grouping diagnostic.
\end{lstlisting}

\clearpage
\subsection*{P6. ORDER BY scope repair}
For contexts requiring ordering expressions to be exposed in the output, this block follows \texttt{expose\_order}. Its candidates refer to the existing projection or its aliases, and the projection~$\Pi$ is preserved.
\begin{lstlisting}[style=prompt]
# Ordering diagnostic
Dialect and query context: <ORDERING_REQUIREMENT_CONTEXT>
Scope: <TARGET_QUERY_SCOPE>
Reported ORDER BY expression: <INVALID_ORDERING_EXPRESSION>

# Existing projection and aliases
<CURRENT_PROJECTED_EXPRESSIONS_AND_ALIASES>

# Admissible ordering references
<CANDIDATES_USING_EXISTING_PROJECTED_EXPRESSIONS_OR_ALIASES>

# Required edit
Rewrite <INVALID_ORDERING_EXPRESSION> using <SELECTED_CANDIDATE>.
Use only the ordering and alias edits authorized by the plan.

# Preserve
Keep all projected expressions and the visible output schema.
Retain the original sort direction and LIMIT.
Preserve the query sources, joins, filters, and aggregation.
Do not add a SELECT expression or remove DISTINCT to hide the error.
\end{lstlisting}

\subsection*{P7. Nested aggregate lifting}
This block describes the \texttt{lift\_agg} rewrite into a CTE or subquery. Original tables and filters may move into the derived scope, but their relational support must be retained as required by the operator contract.
\begin{lstlisting}[style=prompt]
# Nested-aggregate diagnostic
Scope: <TARGET_QUERY_SCOPE>
Invalid expression: <NESTED_AGGREGATE_EXPRESSION>
Inner aggregate: <INNER_AGGREGATE_TO_LIFT>
Outer aggregate: <OUTER_AGGREGATE_TO_RETAIN>

# Supplied lifting plan
Derived scope: <CTE_OR_SUBQUERY_NAME>
Grouping and exported keys: <GROUPING_AND_EXPORT_KEYS>
Exported aggregate alias: <INNER_RESULT_ALIAS>
Permitted CQS edits: <AUTHORIZED_FIELDS_AND_REWRITE>

# Required edit
Compute <INNER_AGGREGATE_TO_LIFT> in the supplied derived scope.
Export its result as <INNER_RESULT_ALIAS>.
Use that result in the original outer aggregate.

# Preserve
Retain the original tables, joins, and filters in the appropriate
scope. Keep their column dependencies and literal values.
Preserve the visible output schema and all non-target components.
Do not discard the outer aggregate or replace it with the inner one.
Use only functions supported by <SQL_DIALECT>.
\end{lstlisting}

\clearpage
\subsection*{P8. Retry after monitor rejection}
The failed checks are passed back as \textit{rej}. The current SQL and resolved-violation history stay unchanged after rejection. Only the relevant feedback fields are included in the next request.
\begin{lstlisting}[style=prompt]
# Current SQL -- the rejected candidate was not applied
<CURRENT_ACCEPTED_SQL>

# Previous verification result
Your last repair was REJECTED.
Failed checks: <FAILED_MONITOR_CHECKS>
Reason: <CONCRETE_FAILURE_DIAGNOSTICS>
Edit that caused rejection: <REJECTED_EDIT_SUMMARY>

# Guidance for this retry
Start from the current SQL above.
Keep <PROTECTED_CLAUSES_EXPRESSIONS_AND_VALUES>.
Do not reintroduce <RESOLVED_VIOLATIONS_IDENTIFIED_IN_FEEDBACK>.
Use only <ADMISSIBLE_REPAIR_CANDIDATES>.
Stay within <PERMITTED_EDIT_FIELDS>.
Correct the reported issue without repeating the rejected edit.

Output one complete corrected SQL query in a SQL code block.
\end{lstlisting}

\subsection*{P9. Multiple diagnostics in one request}
The common request can contain several diagnostics and compatible candidate operators. The prompt supplies the permitted edits for each target; the monitor checks the resulting candidate as a whole.
\begin{lstlisting}[style=prompt]
# Current diagnostics
1. <VIOLATION_1_AND_TARGET_SCOPE>
2. <VIOLATION_2_AND_TARGET_SCOPE>
<ADDITIONAL_DIAGNOSTICS_IF_INCLUDED>

# Admissible repair plan
For diagnostic 1: <OPERATOR_1_AND_CANDIDATES>
For diagnostic 2: <OPERATOR_2_AND_CANDIDATES>
Permitted combined edits: <COMBINED_OPERATOR_EDIT_BOUNDARY>

# Task
Repair the reported violations using the supplied candidates.
Combine compatible fixes within the permitted edit boundary.
Preserve <NON_TARGET_QUERY_COMPONENTS>.
Do not reintroduce <PREVIOUSLY_RESOLVED_VIOLATIONS>.
Do not remove an unrelated clause to make a diagnostic disappear.
Output the complete corrected SQL in a SQL code block.
\end{lstlisting}

\end{document}